\documentclass[11pt,a4paper]{article}
\pdfoutput=1
\usepackage{jheppub}

\usepackage{amsmath,amssymb,amsbsy,amsfonts,latexsym,graphicx}
\usepackage{hyperref}
\usepackage{color,array,subfigure}
       \def\e  {\epsilon}

\renewcommand{\d}{\delta}      
\renewcommand{\l}{\lambda}

\newcommand{\nn}{\nonumber}

\allowdisplaybreaks

 \def\d{{\delta}}

 \def\l{{\lambda}}

 \def\p{\partial}

\def\nn{\nonumber}

\title{  
	Phase Transition in JT Gravity and $T\bar{T}$ Deformation
}
\author[a]{Kyung Kiu Kim,}
\author[a]{Jong-Hyun Baek,}
\author[b, c]{Yunseok Seo}

\affiliation[a]{Department of Physics and Astronomy, Sejong University, Seoul 05006, Korea}
\affiliation[b]{School of Physics and Chemistry, Gwangju Institute of Science and Technology, Gwangju 61005, Korea}
\affiliation[c]{College of General Education, Kookmin University, Seoul, 02707, Korea}

\emailAdd{kimkyungkiu@sejong.ac.kr}
\emailAdd{jonghbaek@gmail.com}
\emailAdd{yseo@kookmin.ac.kr}

\abstract{
In this paper we study a black hole phase transition in a generalized JT gravity noticed in 2006.03494. 
We investigate the effect of the phase transition on the Euclidean geodesic and holographic two-point function in models with dilaton potential which interpolates two ordinary JT gravities with different cosmological constants. It is noted that there exists a closed geodesic with a new scale at low temperature phase when the potential has a locally negative region. This scale causes several peaks in the two-point function. We also comment on the phase transition of charged black holes. We then consider coupling generalized JT gravity to a matter and study its relation to a $T\bar{T}$ deformation of CFT at the classical level. We find the deformation parameter as a function of the dilaton and provide examples showing Janus-type couplings.}


\begin{document}
\maketitle
\flushbottom

%%%%%%%%%%%%%%%%%%%
\section{Introduction}
%%%%%%%%%%%%%%%%%%%

Even though two-dimensional gravity is simple, it plays a very important role in theoretical physics. For Jackiw-Teitelboim (JT) gravity \cite{Teitelboim:1983ux, Jackiw:1984je}, it has been shown that a field theory dual is identified with a random ensemble of Sachdev-Ye-Kitaev (SYK) model \cite{Sachdev:1992fk, KitaevTalks} describing maximally chaotic system \cite{Jensen:2016pah,Maldacena:2016upp, Engelsoy:2016xyb}. In other application, it is noted that the information problem, one of the salient problems in physics, can be realized in this simple model \cite{Almheiri:2019hni,Almheiri:2019qdq}. These successful investigations indicate that two-dimensional gravity deserves further study.

Recently, there was an interesting study on a first order phase transition among black holes in a generalized JT gravity with diltaon-potential \cite{Witten:2020ert}. 
%Like the JT gravity \cite{Saad:2019lba}, this theory has also a dual description of matrix model \cite{Witten:2020wvy}. Such a study indicates a possibility of the first order phase transition encoded in the density of state for the matrix model. This supports the first order phase transition can happen in the gravity model. 
In our work, we revisit this phase transition in specific models described by diltaon-potentials, which connects two JT gravities with different negative cosmological constants. The origin of the phase transition comes from the fact that there exists a certain region in temperature, where the black hole geometry is not determined uniquely. Thus, one has to compare free energies of all possible black holes to find the most preferable black hole for a given temperature. As the temperature varies, the bulk geometry undergoes a radical change in entropy and the location of horizon. We show this change explicitly in our specific JT gravity models.

In addition to this sudden change under the phase transition, we found that the low temperature phase has a significantly different characteristic if we allow a locally negative region of the potential. Since the metric function is given by the integration of the potential after choosing a suitable ansatz, a negative region of the potential produces a local minimum of the metric component. The location of this local minimum plays a role of a cut-off for Euclidean geodesics, which are anchored to the boundary. This effective cut-off introduces a scale in the bulk as well as boundary quantum mechanics, which is different from the temperature. Naturally, the scale is given by the length of a closed geodesic surrounding the local minimum of the metric component.

It is also known that holographic Green's functions have a well-defined geodesic limit. Accordingly, one may expect that this scale should appear in the holographic two-point functions. We show that this is indeed the case by a numerical study of scalar two-point functions. The two-point functions exhibit peaks around the scale. This phenomenon can also be understood by the equation of motion for the corresponding bulk field, which looks like a Schr$\ddot{\text{o}}$ding equation.

Another interesting aspect of JT gravity is that its relation to the $T\bar{T}$ deformation \cite{Smirnov:2016lqw, Cavaglia:2016oda} when coupled to a conformal matter. In \cite{Dubovsky:2018bmo}, it is shown that flat JT gravity coupled to matter is equivalent to $T\bar{T}$ deformation of the matter action in flat space by using the first order formalism of gravity. The flat JT gravity has a constant potential, which looks like a cosmological constant, so the resulting deformed action has a constant deformation parameter for the $T\bar{T}$ operator. It is important to note that a dynamical coordinate appears and provides dynamical zweibein in the deformed action. The interpretation of this coordinate has been studied in various contexts. \cite{Dubovsky:2017cnj,Cardy:2018sdv,Conti:2018tca,Conti:2019dxg,Coleman:2019dvf,Aguilera-Damia:2019tpe, Tolley:2019nmm, Mazenc:2019cfg}

In this paper, we extend the derivation to JT gravity with a general diltaon potential. The on-shell gravity action is rewitten as a $T\bar{T}$ deformed action for conformal matter. However, the deformation parameter is not a constant but a function of the dilaton. This dilaton is a function of a radial coordinate in generic case. Thus, the deformation parameter depends on a space coordinate.

As simple examples, we consider some vacuum solutions without horizon. For JT gravity or AP model \cite{Almheiri:2014cka}, the $T\bar{T}$ coupling is still constant, but the base manifold is (Euclidean) $AdS_2$, where the $T\bar{T}$ operator is well-defined as in \cite{Brennan:2020dkw}. 
For JT gravity with a potential which interpolates two different values of the cosmological constants, we consider a domain wall geometry as the base manifold. This can be obtained by the zero temperature limit of the black hole solution. In this case, the $T\bar{T}$ coupling function becomes a local function which approaches to a different value for each asymptotic $AdS_2$.
 For asymptotically flat space, a kink-like potential is considered, for which the base manifold has a simple disc topology but with a flat interior region. The corresponding coupling function also has two plateaus for the flat space regions.

This paper is organized as follows. In section \ref{sec2}, we study phase transitions of black holes in generalized JT gravity which has a potential interpolating between two JT gravities of different cosmological constants. The effects of phase transitions on the two-point function and the geodesic are investigated. In section \ref{sec3}, charged black holes are analyzed using similar methods. In section \ref{sec4}, a connection of generalized JT gravity to $T\bar{T}$ deformation is explained.

%%%%%%%%%%%%%%%%
\section{Black Holes in Interpolating JT Gravities}\label{sec2} 
%%%%%%%%%%%%%%

In this section we investigate black hole phase transitions discussed in \cite{Witten:2020ert}. In order to find the physical implication of the phase transition, we consider Euclidean geodesics and holographic two-point function in specific models. 
%Also, we extend the two-dimensional gravity model with a Maxwell field and explain how this extension affects the diltaon-potential effectively.  

\subsection{Black hole phase transition in two dimensions}\label{section21}

We start with a brief summary of the phase transition studied in \cite{Witten:2020ert} using our two dimensional gravity models. The Euclidean action is given by
\begin{align}\label{Sb}
S_{GJT} =  - \frac{1}{2} \int_{\mathcal{M}} d^2 x \sqrt{g} \big[\phi_0 R + \phi R + W(\phi) \big] - \sum_{\mathcal{I}}\int_{(\partial\mathcal{M})_{\mathcal{I}}} d\tau \sqrt{\gamma}\big[(\phi_0+\phi) K - \mathcal{L}^c_{\mathcal{I}} \big],
\end{align}
where $\phi_0$ is a constant, so the first term is just the Einstein-Hilbert action, which is a total derivative in two dimensions. Nevertherless, this term contributes to black hole entropy, as explained in \cite{Witten:2020ert}, and it is important in the computation of path integrals. The Gibbons-Hawking term proportional to $K$ guarantees a well-defined Dirichlet problem for the metric and $\gamma$ is the induced einbein resulting from the ADM decomposition. In addition $\mathcal{L}^c_{\mathcal I}$'s are the counter terms, introduced to make the on-shell action finite. The index $\mathcal{I}$ stands for the boundaries of the two-dimensional manifold $\mathcal{M}$. For JT gravity, the diltaon-potential is given by $W=2 \phi$ with the unit AdS radius. The corresponding counter term is given by $\mathcal{L}^c_{\mathcal{I}} = \phi$.

Now we take an ansatz as follows:
\begin{align}
ds^2 = A(x) d\tau^2 + \frac{dx^2}{G(x)}~,~\phi = \phi(x),
\end{align}
where $G(x)$ can be taken as $A(x)$  by a coordinate transformation. Using this metric ansatz without fixing $G(x)$, the bulk action can be written as follows:
\begin{align}
\mathcal{L}_b =\frac{\sqrt{G} \phi A''}{2 \sqrt{A}}+\frac{\phi A' G'}{4 \sqrt{A} \sqrt{G}}-\frac{\sqrt{G} \phi A'^2}{4 A^{3/2}}-\frac{\sqrt{A} W(\phi)}{2 \sqrt{G}}~.
\end{align}
The equations of motion with fixing gauge as $G(x)=A(x)$ lead to 
\begin{align}
\partial_\phi W(\phi)-A''(x)=0~,~W(\phi)-A'(x)\phi'(x)=0~.
\end{align}
In addition the equation of motion for the dilaton is nothing but $\phi''(x)=0$. The solution is simply given by $\phi(x) = C_0 + C_1 x$. After a coordinate transformation $x \to \frac{x-C_0}{C_1}$, the dilaton $\phi$ can be chosen as $\phi(x)=x$. From now on, we identify the coordinate $x$ with $\phi$. Then, the only equation we have to solve is
\begin{align}
W(x) = A'(x)~.
\end{align} 
Therefore, the general solution for metric can be written as 
\begin{align}\label{A0}
A(x) = \int_{x_0}^x dx' W(x') + A(x_0)~,
\end{align}
where $x_0$ is a certain position in the radial coordinate. We would like to consider cases with only positive $A(x)$ for a well-defined Euclidean geometry.

\begin{figure}[t] 
\begin{centering}
    \subfigure[ ]
    {\includegraphics[width=9.1cm]{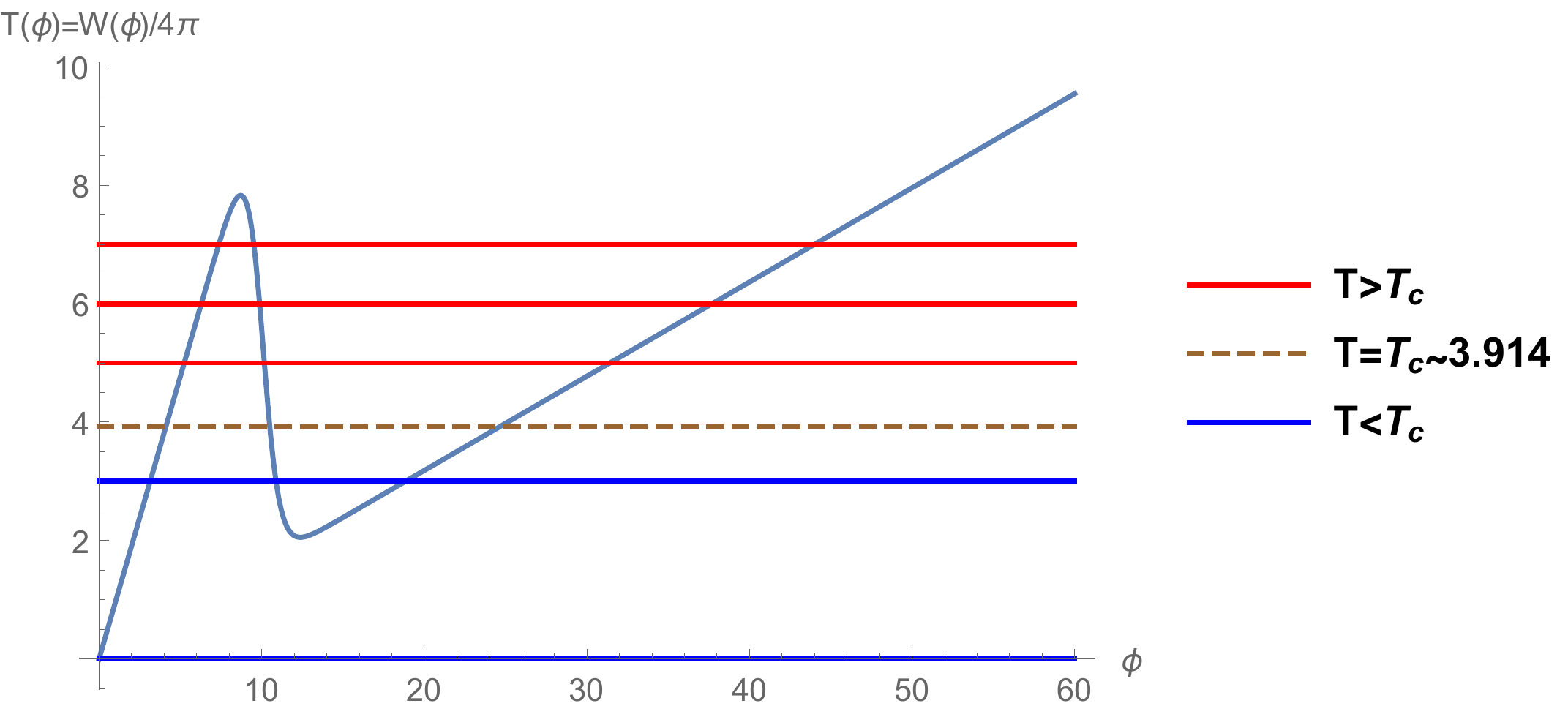}  }
        \subfigure[ ]
    {\includegraphics[width=5.6cm]{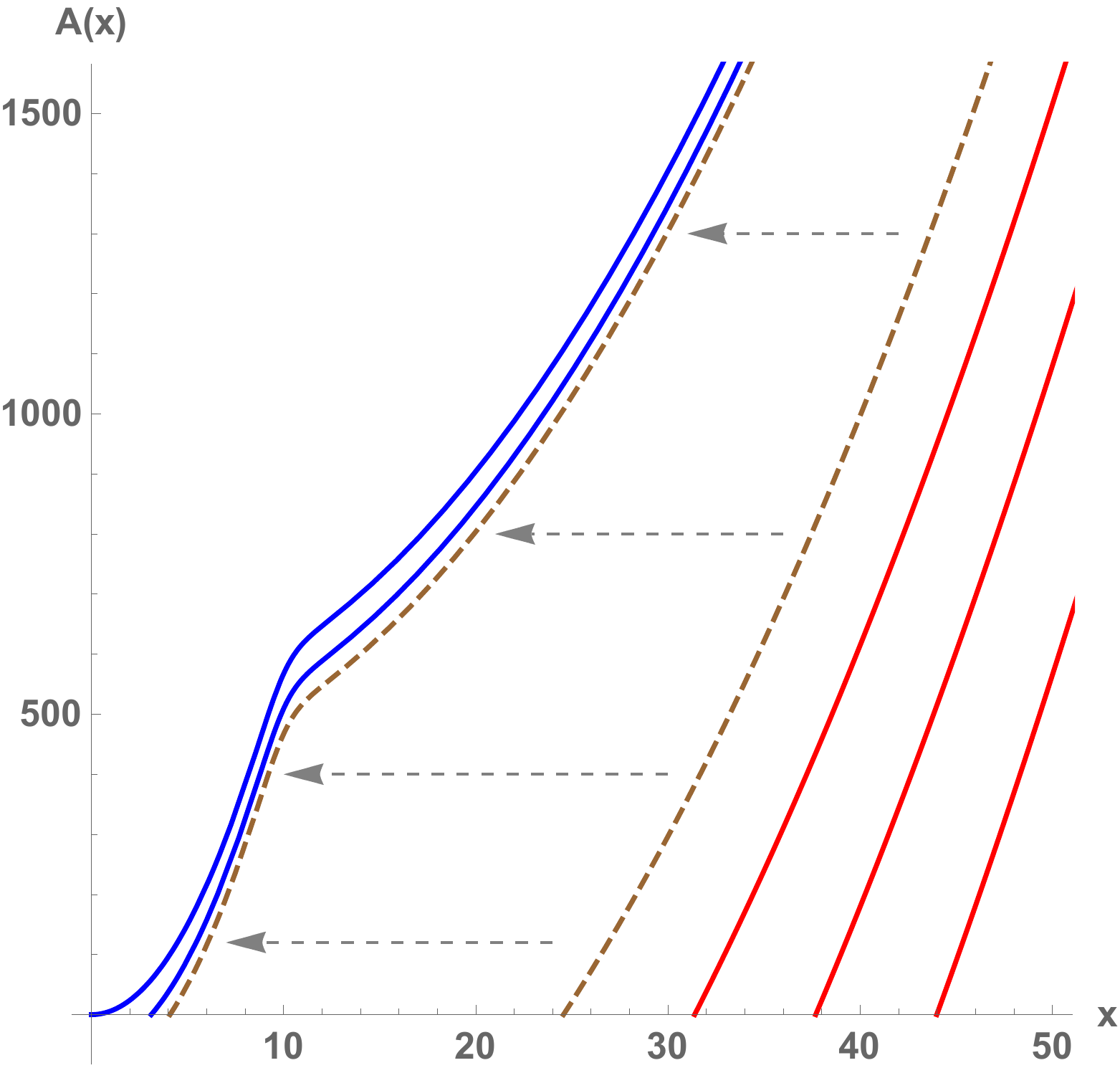}  }
       
    \caption{(a) shows the potential (\ref{potential01}) and temperature lines. Intersection points with the potential and the temperature lines are possible horizon locations for a given temperature. Except for the lowest temperature line, the  intersection points in the middle are excluded since they give negative heat capacity. The brown dashed line denotes the critical temperature of the phase transition. (b) shows the corresponding metric function $A(x)$. The left and right dashed-brown curves depict geometries just below and above $T_c$, respectively.
} \label{fig:Tphi}
\end{centering}
\end{figure}

In \cite{Witten:2020ert}, it was shown that a phase transition may occur among asymptotically AdS black holes when the equation $W(\phi) = 4\pi {T}$ has more than three roots for a positive constant temperature $T$. To explain this phase transition more explicitly, let us take a specific form of the diltaon-potential as follows:
\begin{align}\label{potential01}
W(\phi) = 4\pi {T}(\phi) =  2 \phi +5 \phi  \big[\tanh (\phi +10)-\tanh (\phi-10 )\big],
\end{align} 
where ${T}(\phi)$ is just a potential scaled by $4\pi$.\footnote{ {We distinguish between $T$ and $T(\phi)$} as in \cite{Witten:2020ert}.} This is plotted in Figure \ref{fig:Tphi}. This potential becomes the JT gravity $W(\phi)\sim 2\phi$ for large $\phi$. On the other hand, the potential becomes another JT gravity with a different cosmological constant ($W(\phi)\sim 12 \phi$) in the small $\phi$ region. Therefore, one may regard this specific potential as describing a model which interpolates two JT gravities.

Now, we consider black hole geometries. The temperature of a black hole is given by $A'(x_h)/4\pi$, where $x_h$ is the location of the horizon or a dilaton value $\phi_h$. So it can be written in terms of the potential as follows:
\begin{align}
T \equiv T(\phi_h)= \frac{1}{4\pi}W(x_h)~.
\end{align}
Since a given temperature determines possible locations of horizon, the integration constant in (\ref{A0}) can be fixed and the metric function becomes just the integration of the potential
\begin{align}
A(x) = \int_{x_h}^x dx' W(x')~.
\end{align}

 As one can see in Figure \ref{fig:Tphi} (a), there are three candidates for possible horizon location in certain range of temperature. They are given by the intersection points of the temperature lines and the dilaton-potential except for the lowest temperature line. Among these candidates, the middle locations can be excluded because the corresponding black hole geometries have negative heat capacities. For the other possible locations, one needs to compare the free energies of the corresponding black holes. As shown in \cite{Witten:2020ert}, the free energy difference between two black holes is
\begin{align}\label{dF}
\Delta F = F_R - F_L = 2\pi \int_{\phi_L}^{\phi_R} d\phi \left( T(\phi)-T \right)~,
\end{align}
where $F_L$ and $F_R$ are the free energies of the black holes which have the left and right intersecting points, $\phi_L$ and $\phi_R$ for a given temperature. Therefore, the black hole with the horizon $\phi_R$ is dominant in the high temperature and the black hole with the horizon $\phi_L$ is preferable in the low temperature. There exists a critical temperature $T_c$ satisfying $\int_{\phi_L}^{\phi_R} d\phi \left( T(\phi)-T_c \right)=0$.

This phase transition gives rise to a drastic change in geometry. It also causes an entropy jump given by $\Delta S = 4\pi \left( \phi_R-\phi_L \right)$. They can be seen in Figure \ref{fig:Tphi}(b), which shows the metric function $A(x)$. The red and blue curves correspond to high and low temperature black holes, respectively. The brown-dashed curves denote geometries just above and below the critical temperature. The arrows indicate the change from high to low temperature.

 \begin{figure}[t] 
 	\begin{centering}
 		\subfigure[ ]
 		{\includegraphics[width=9.1cm]{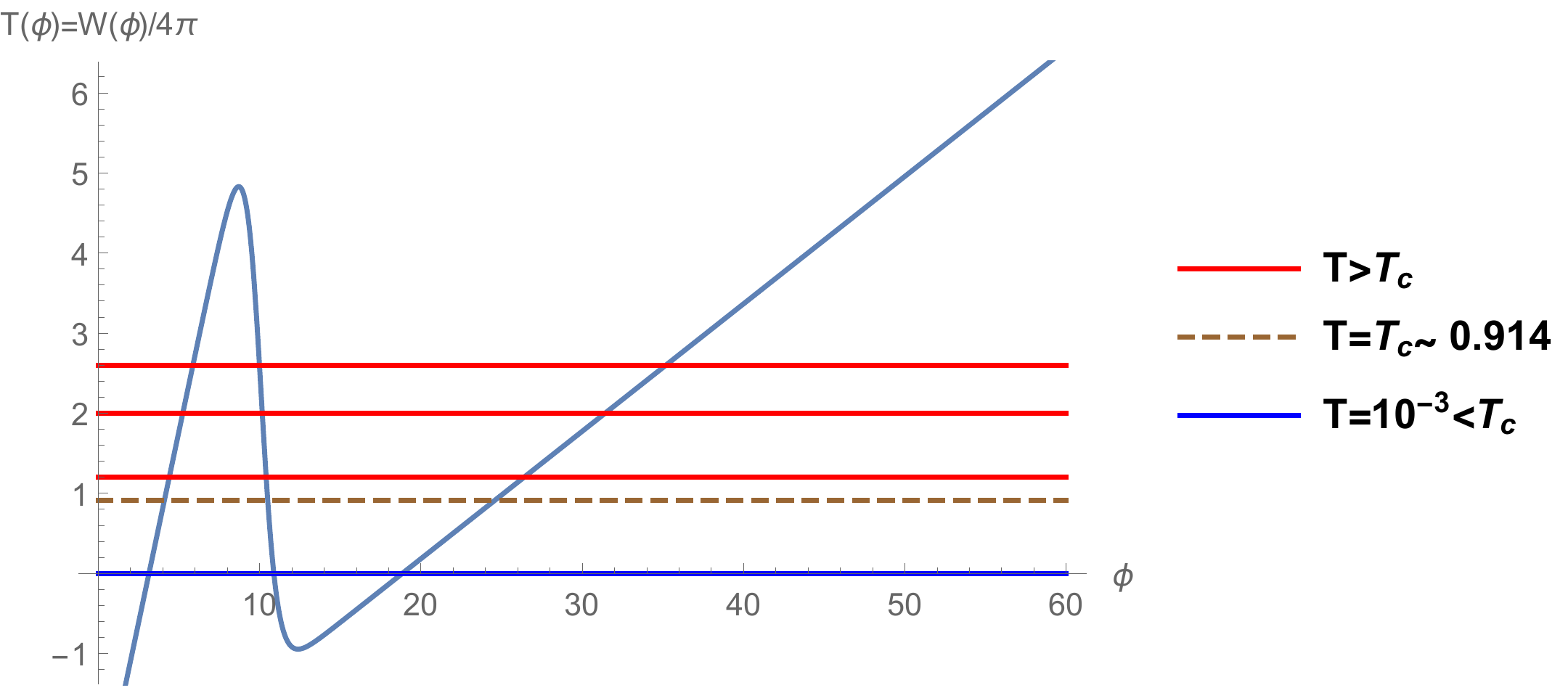}  }
 		\subfigure[ ]
 		{\includegraphics[width=5.6cm]{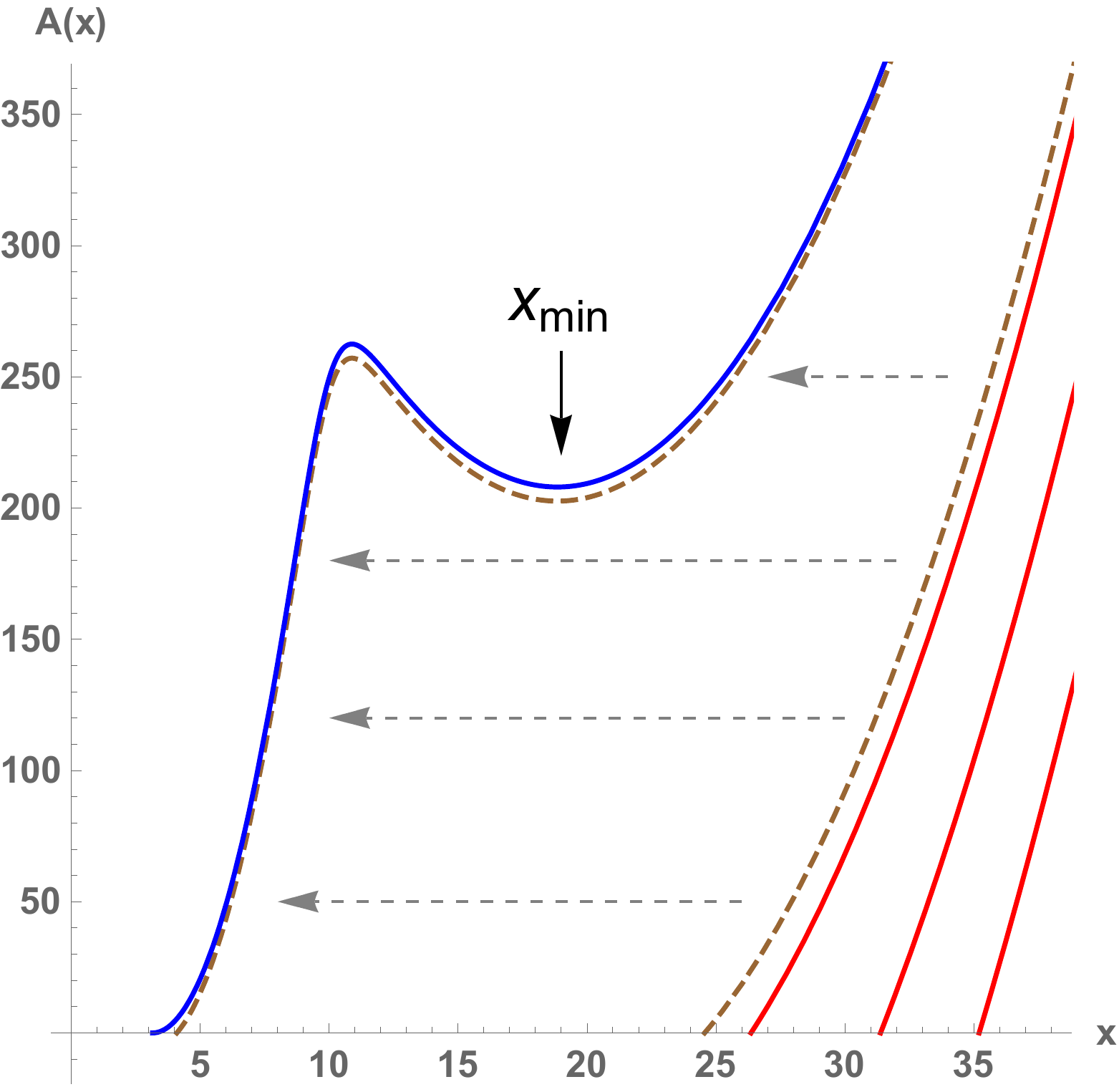}  }
 		
 		\caption{(a) shows a locally negative potential \eqref{Wnegative} and temperature lines. The critical temperature is given by $T_c\sim 5-\frac{45}{4 \pi}$, which is denoted by the brown dashed line. (b) shows the metric function $A(x)$. The blue and red curves stand for low and high temperature black holes, respectively. The brown curves depict geometries just below and above $T_c$. Below the critical temperature, geometries have local minima at $x_{min}$.
 		} \label{fig:Negative_W}
 	\end{centering}
 \end{figure}

Now, we introduce a more interesting case with a potential which has a negative region. This case can easily be achieved by subtracting a constant to the previous potential, i.e, $W(\phi)\to W(\phi)-\mathcal{C}$. In the following, we take $\mathcal{C}=12 \pi$. Namely,
\begin{align}\label{Wnegative}
W(\phi) = 4\pi T(\phi) =  2 \phi +5 \phi  \big[\tanh (\phi +10)-\tanh (\phi-10 )\big]-12\pi.
\end{align}
The potential and the corresponding metric are plotted in Figure \ref{fig:Negative_W}. Here one may notice that $A(x)$ below the critical temperature has a local minimum $x_{min}$ due to the negative part of the potential. 

This qualitative difference in geometry results in a significant consequence. In order to see the effect of the existence of local minimum, we will study Euclidean geodesics and holographic two-point functions in the following subsections.

%%%%%%%%%%%%%%%%%%%%%%
\subsection{Euclidean geodesic}
%%%%%%%%%%%%%%%%%%%%%%

As a first observation on physics, we take geodesics in the black hole geometry. Since we are interested in either asymptotically AdS geometry or the potential which approaches to that of JT gravity for large $\phi$, we assume that there is a dual boundary quantum mechanics to the black hole. We start with analysis for geodesics anchored to the boundary of  AdS spacetime. The geodesic $x(\tau)$ has the boundary condition $x(\pm l/2)=\infty$, where $l$ is the distance between the starting and end points at the boundary.

Then, the induced einbein for a geodesic is 
\begin{align}
ds_{ind}^2 = \left( A(x(\tau)) + \frac{x'^2}{A(x(\tau))} \right) d\tau^2~.
\end{align}
The geodesic length is given by
\begin{align}
I_g(l) = 2 \int_0^{l/2} d\tau \sqrt{A(x(\tau)) + \frac{x'^2}{A(x(\tau))}},
\end{align}
where the effective Lagrangian has no explicit dependence of $\tau$, so one can find the conserved Hamiltonian given by
\begin{align}\label{HforGeodesic}
\mathcal{H} = -\frac{A(x(\tau ))}{\sqrt{\frac{x'(\tau )^2}{A(x(\tau ))}+A(x(\tau ))}}~.
\end{align}
At the tip ($\tau=0$) of a geodesic, we impose $x'(0)=0$ for a regular curve. So the Hamiltonian $\mathcal{H}$ is nothing but $\mathcal{H}=-\sqrt{A_*}$, where $A_* = A(x(0))$. Using the expression of $x'(\tau)$ from (\ref{HforGeodesic}),\footnote{$x'=\frac{A(x) \sqrt{A(x)-A_*}}{\sqrt{A_*}}$} the geodesic length becomes
\begin{align}\label{geodesic}
I_g(l) = 2 \int_{x_*}^{\Lambda_c} dx \frac{1}{\sqrt{A(x)-A_*}}~,
\end{align}
where $\Lambda_c$ is the UV cut-off and $x_*$ is the position at the tip of geodesic, i.e. $x(\tau=0)$. Near the boundary, this length has a logarithmic divergence. This is reminiscent of the entanglement entropy for 1+1 dimensional field theories.

Now, we apply this formula to the potential with a negative region plotted in Figure \ref{fig:Negative_W} (a). In the low temperature below the $T_c$, $A(x)$ has a local minimum $A_{min}=A(x_{min})$.  So (\ref{geodesic}) is not well-defined for $x_*<x_{min}$. This implies that the Euclidean geodesics starting from boundary cannot cross the $x=x_{min}$ point.  When $x_*$, which is larger than $x_{min}$, approaches $x_{min}$, one may expect the boundary points of the geodesic approach certain value $x=\pm l_p/2$. For $l$ slightly greater than $l_p$, the tip $x_*$ of geodesic is still very near $x_{min}$ but the geodesic is deformed and its length is approximated to the value, $I_g (l>l_p)=  \lim_{x_*\to x_{min}}\int_{x_*}^{\Lambda_c} dx \frac{2}{\sqrt{A(x)-A(x_*)}}$. We are interested in geodesics related to the two-point function, so we consider geodesics which exist outside of $x=x_{min}$.\footnote{For $x_*<x_{min}$, i.e., a geodesic starting from $x=x_*$ and $\tau=0$ case, the formula $I_g(l) =  \int_{x_*} dx \frac{2}{\sqrt{A(x)-A_*}}$ is still valid. Now, however, corresponding geodesics head for the black hole horizon. Interestingly, there is a set of maximal closed geodesics containing the geodesics wrapping $x=x_{max}$, where $x_{max}$ is the location which gives the local maximum of $A(x)$.  It would be interesting to study the physical implication of this set of closed geodesics. } Therefore, for $l>l_p$, the minimum value of $A_*$ can be replaced with $A_{min}$ in (\ref{geodesic}). 
In fact, the period of the time circle is important for this speculation because $l_p$ should be smaller than $\beta/2 = 1/2T$. For a very low temperature, at which phase transition can occur, this is the case.
	
There exists another type of geodesic, which is a closed minimal geodesic wrapping $x=x_{min}$. It satisfies the minimal geodesic equation, $x' = \frac{A(x)\sqrt{A(x)-A_*}}{\sqrt{A_*}}$. This gives rise to a new scale which is different from the temperature. It can be easily computed as
\begin{align}\label{scale00}
\Lambda_{s}=\int_0^{1/T} \sqrt{A(x_{min})}d\tau = \frac{\sqrt{A_{min}}}{T}~. 
\end{align} 
It is desirable to see how this scale can appear in other physical quantities. In fact, a geodesic anchored to the boundary are closely related to the two-point function in the boundary theory. Thus, we turn to two-point functions in the next subsection.

%%%%%%%%%%%%%%%%%%%%
\subsection{Holographic two-point function}
%%%%%%%%%%%%%%%%%%%%%

In this subsection, we focus on a boundary two-point function using a holographic method. We will consider an in-going fluctuation mode and use the Lorentzian signature for the metric 
\begin{align}
ds^2 = - A(r) dt^2 + \frac{1}{A(r)} dr^2~,
\end{align} 
where we use $t$ and $r$ as time and radial coordinates instead of $\tau$ and $x$ of the Euclidean spacetime. As an additional probe field in this background, we consider a scalar field. The total action is given by
\begin{align}
S_t = S_{GJT} +\int dx^2 \sqrt{-g} \left( - \frac{1}{2} (\nabla \psi)^2 -\frac{1}{2}m^2 \psi \right) + S^c_{\psi} ~,
\end{align}
where $S_{GJT}$ is the generalized JT gravity (\ref{Sb}) with the Lorentzian signature and $S^c_\psi$ is the suitable counter term action for this scalar field.

The equation of motion of the scalar for a frequency $\omega$ is given by 
\begin{align}\label{scalarEq02}
\varphi ''(r)+\frac{A'(r) \varphi '(r)}{A(r)}-\frac{m^2 \varphi (r)}{A(r)}+\frac{\omega ^2 \varphi (r)}{A(r)^2}=0~,
\end{align}
where we have used $\psi = \varphi(r)e^{-i\omega t}$. Near the boundary of the geometry, the asymptotic solution with vanishing $\omega$ is
\begin{align}
\varphi(r) \sim   r^{-\frac{1}{2}\pm \sqrt{m^2+\frac{1}{4}}   }.
\end{align}
This indicates that the BF bound is given by $m^2= -\frac{1}{4}$. For a special case of $m^2=0$, the asymptotic solutions are a constant and $1/r$. The constant solution is a non-normalizable mode and $\frac{1}{r}$ corresponds to a normalizable mode. By following the standard AdS/CFT prescription, the coefficient of the two modes can be identified with the expectation value of a dimension one ($\Delta=1$) operator and its source. In general, the asymptotic behavior of the field is given by
\begin{align}\label{AsympExpansion}
\psi \sim \left(J(\omega)\, r^{-\frac{1}{2}+ \sqrt{m^2+\frac{1}{4}}}+\cdots + \mathcal{O}(\omega)\,   r^{-\frac{1}{2}- \sqrt{m^2+\frac{1}{4}}} + \cdots\right) e^{-i\omega t}~,
\end{align}
where $J(\omega)$ and $\mathcal{O}(\omega)$ are the Fourier mode of source and expectation value for a dimension $\Delta =\frac{1}{2}+\sqrt{m^2 + \frac{1}{4}}$ operator.

Now, let us consider the two-point function of scalar operators. To obtain this quantity by a numerical calculation, we impose the in-going boundary condition at the horizon. The metric function $A(r)$ can be expanded near horizon as
\begin{align}
A(r) \sim (r-r_{H}) A'(r_H) + \cdots = (r-r_H) W(r_H)+ \cdots.
\end{align}
If we assume  $\varphi (r) \sim (r-r_H)^{\alpha}$ at the horizon, leading expression of (\ref{scalarEq02}) becomes
\begin{align}
\left( \alpha^2 +\frac{\omega^2}{W(r_H)^2}\right) (r-r_H)^{(\alpha-2)} =0,
\end{align}
and the solution is $\alpha =\pm i \frac{\omega}{W(r_H)}$. The positive (or negative) sign corresponds to outgoing(or infalling) wave in tortoise coordinate. The natural choice for the boundary condition at horizon is the infalling wave. Taking into account $e^{-i \omega t}$, we choose the following condition at horizon:
\begin{align}\label{in-going}
\psi \sim e^{-i\omega \left(t + \log(r-r_h)/W(r_h) \right)}~.
\end{align}

\begin{figure}[t] 
\begin{centering}
    \subfigure[ ]
    {\includegraphics[width=7cm]{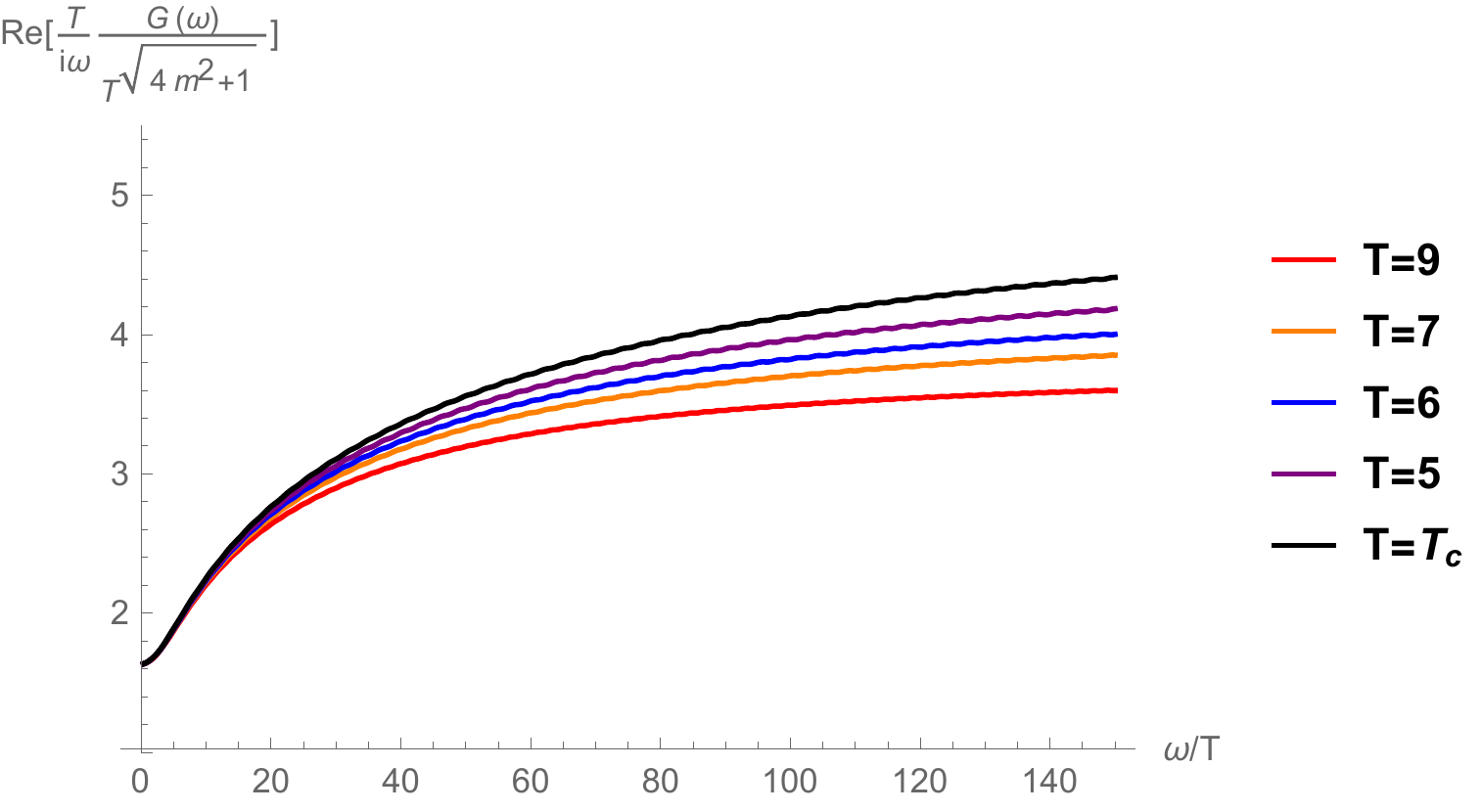}  }
       \subfigure[ ]
   {\includegraphics[width=7cm]{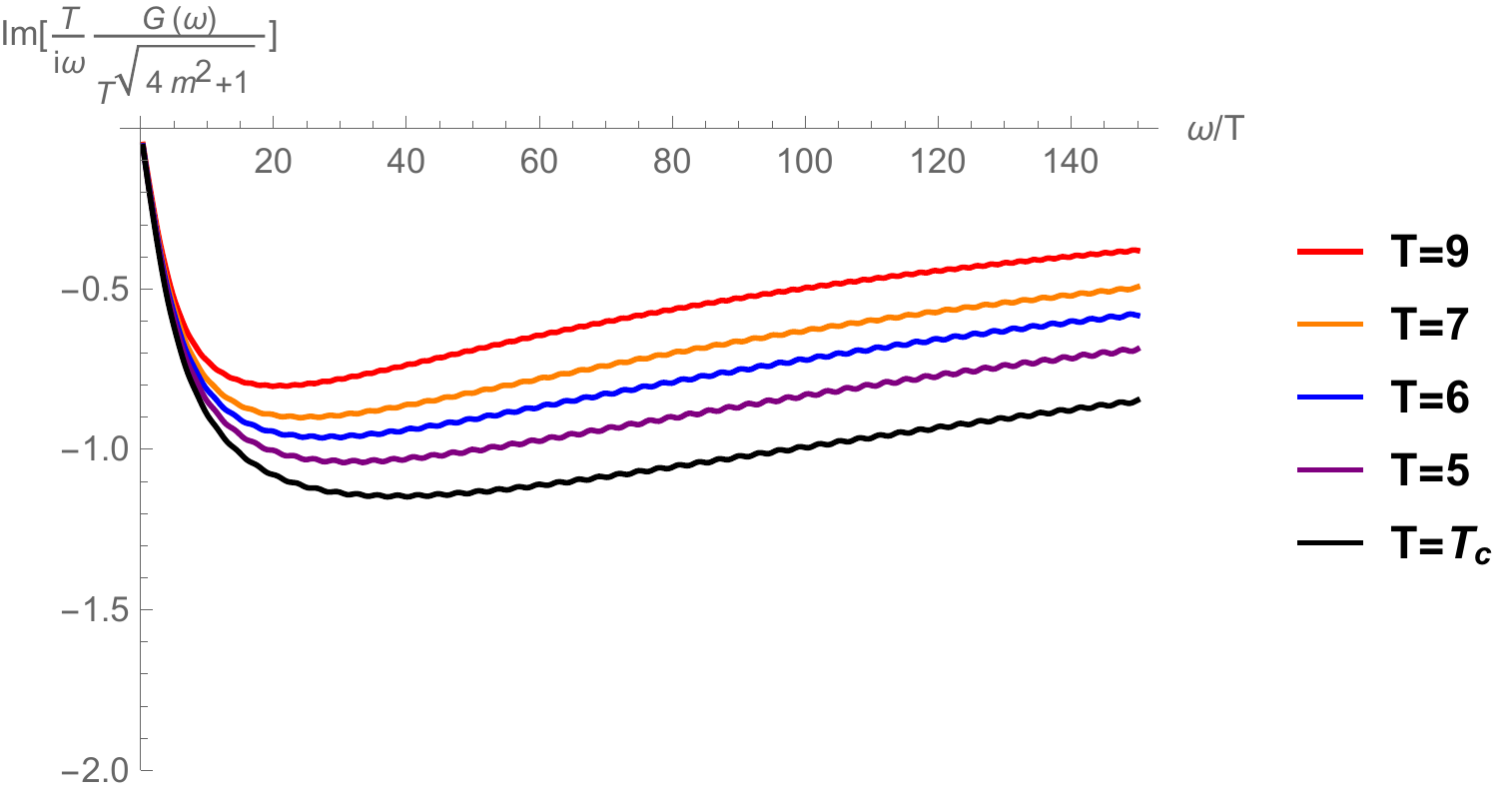}  }\\
     \subfigure[ ]
    {\includegraphics[width=7cm]{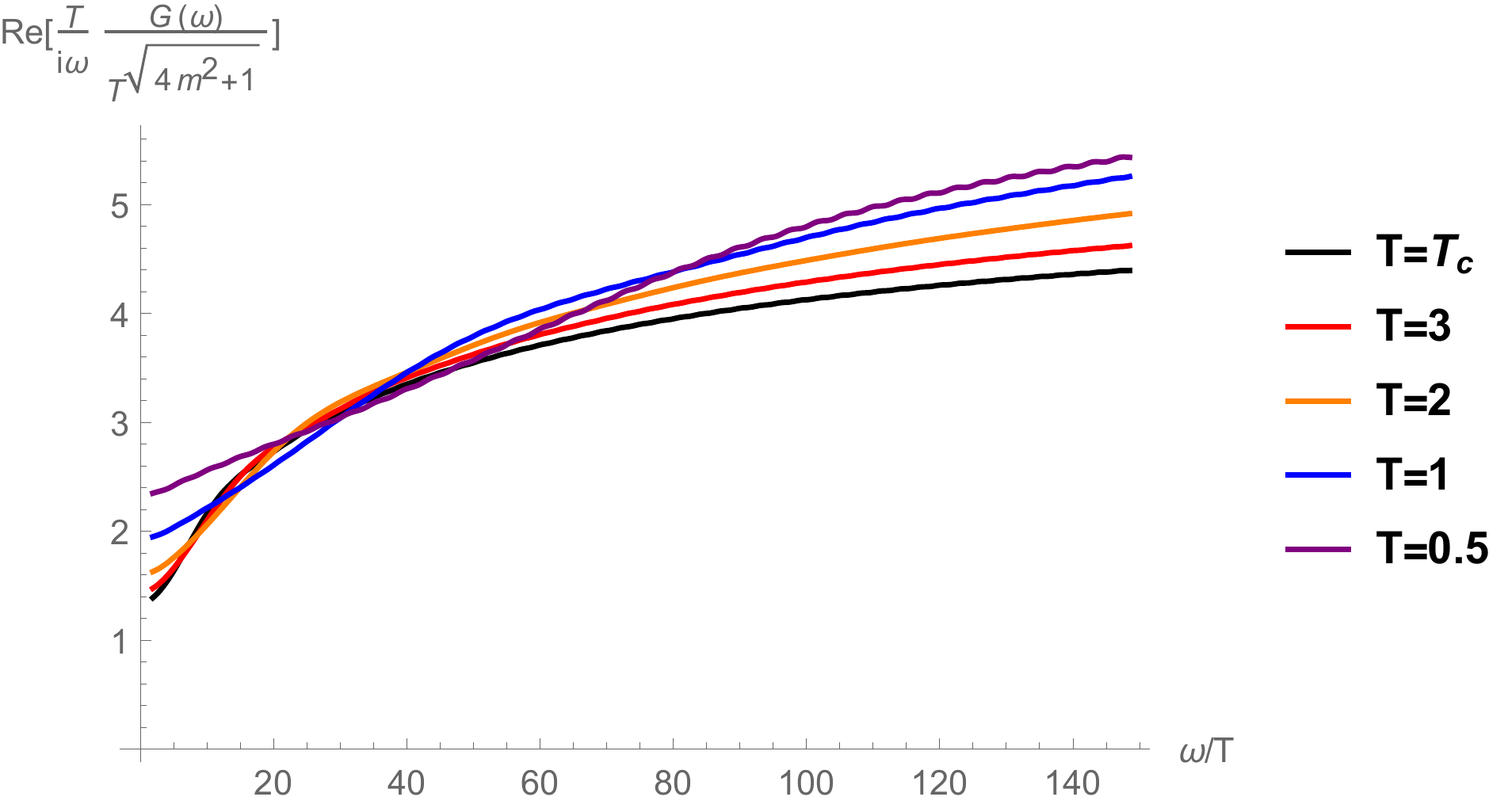}  }
       \subfigure[ ]
   {\includegraphics[width=7cm]{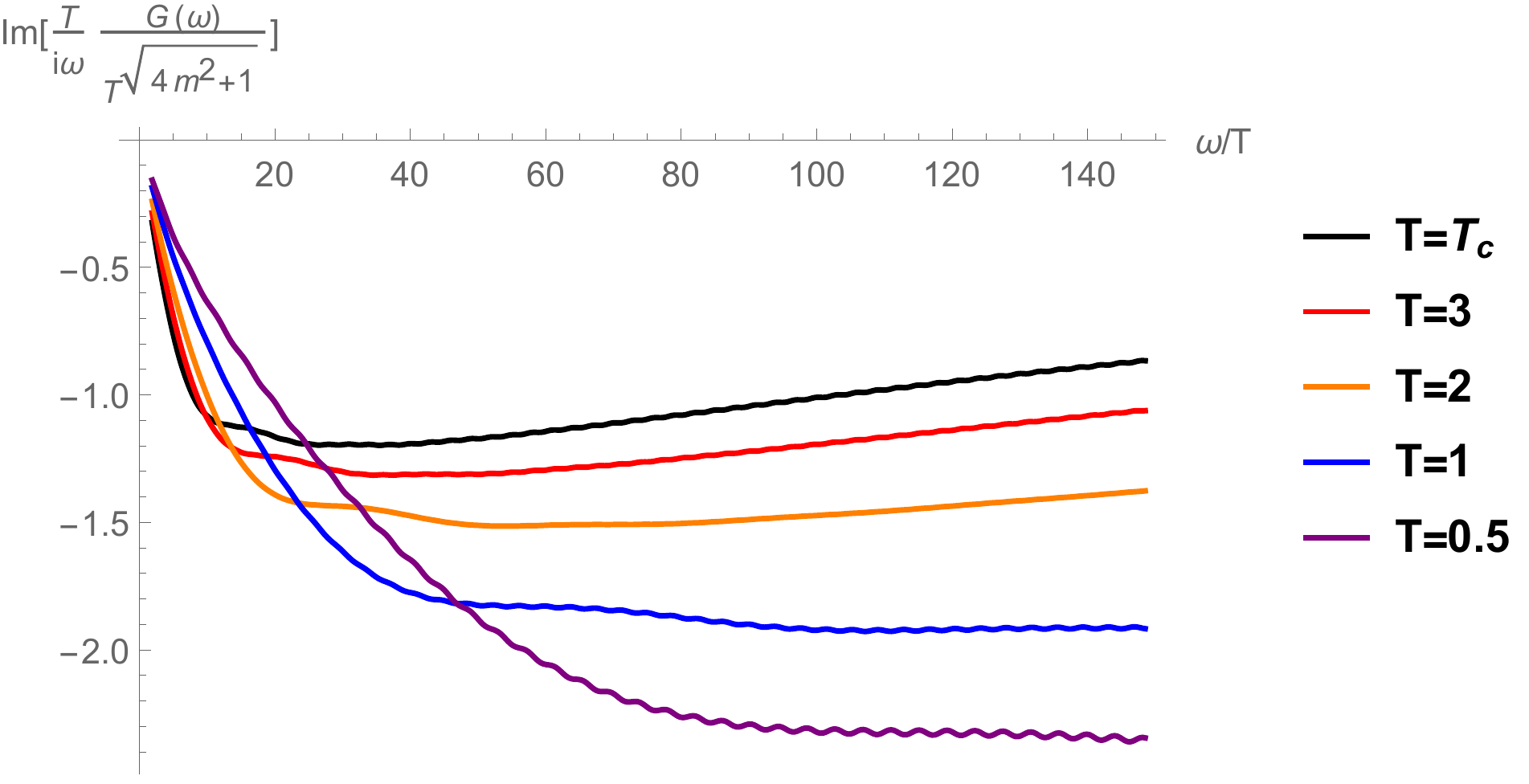}  }   
    \caption{ These figures show the two-point function with $m^2=0.2$. (c) and (d) are obtained by a curve fitting and the error $\mathbb{E}$ of the curves is given by $\mathbb{E}<4\times 10^{-2}$.
} \label{fig:Positive_m_02}
\end{centering}
\end{figure}

\begin{figure}[t] 
\begin{centering}
    \subfigure[ ]
    {\includegraphics[width=7cm]{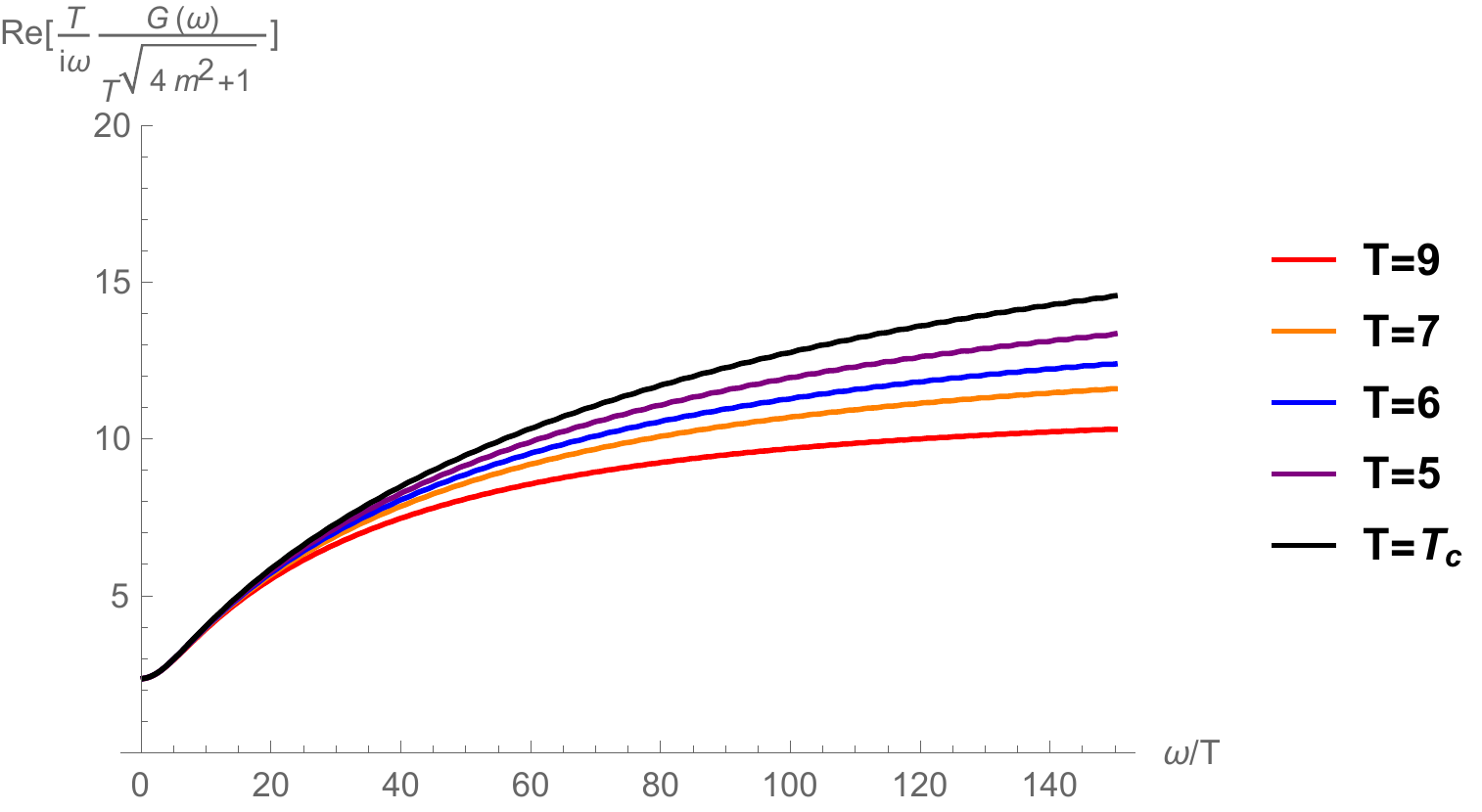}  }
       \subfigure[ ]
   {\includegraphics[width=7cm]{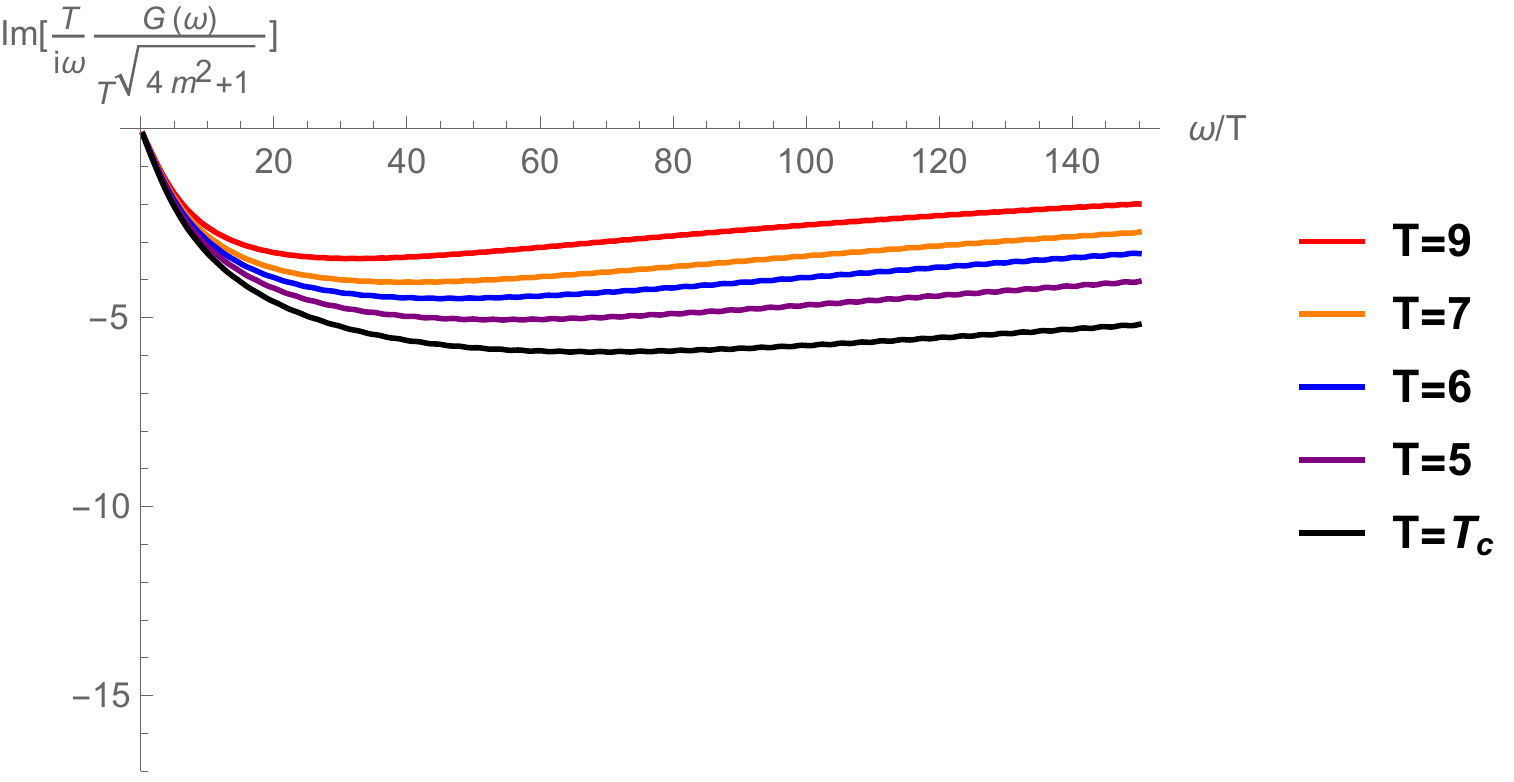}  }\\
     \subfigure[ ]
    {\includegraphics[width=7cm]{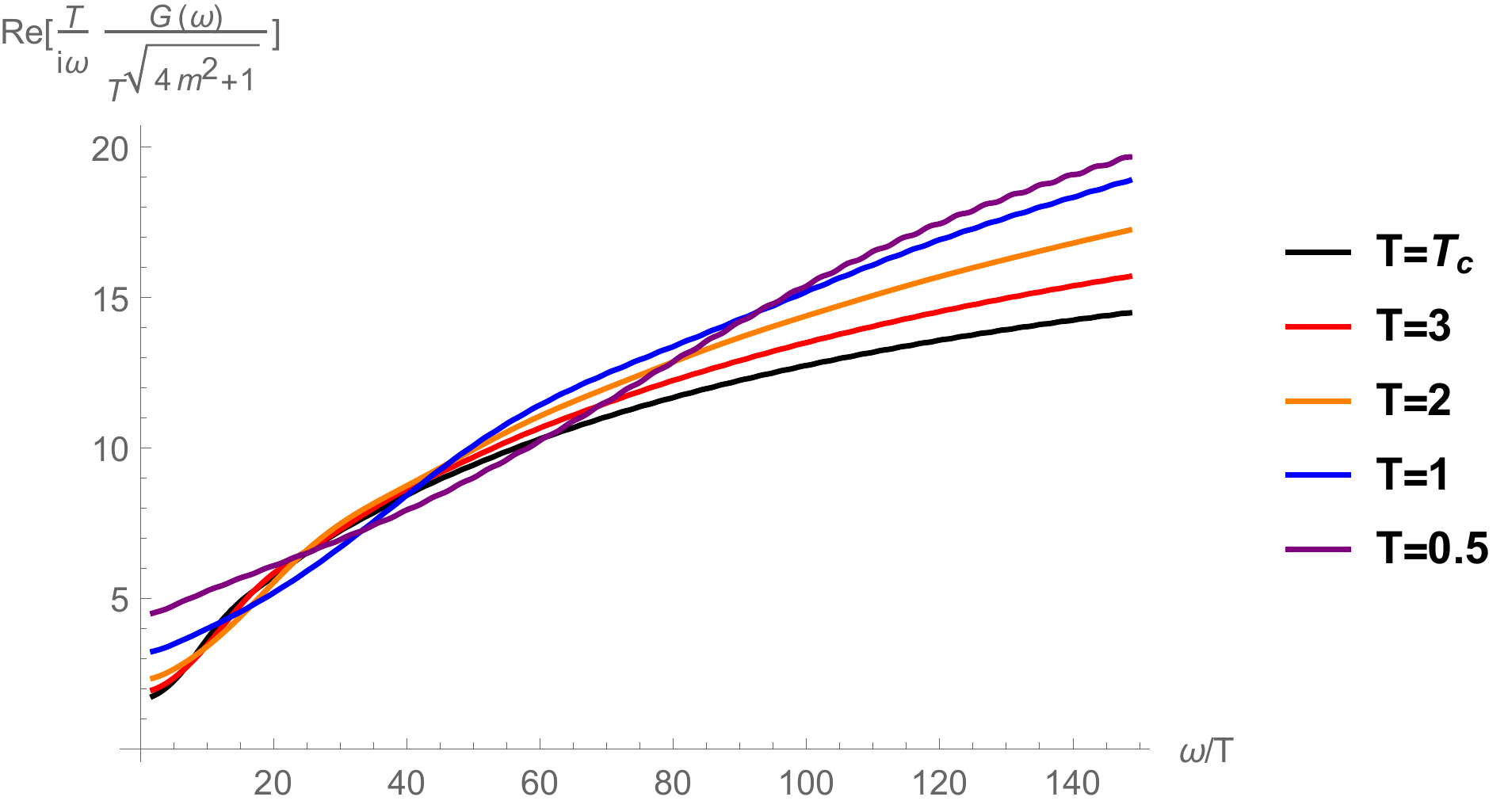}  }
       \subfigure[ ]
   {\includegraphics[width=7cm]{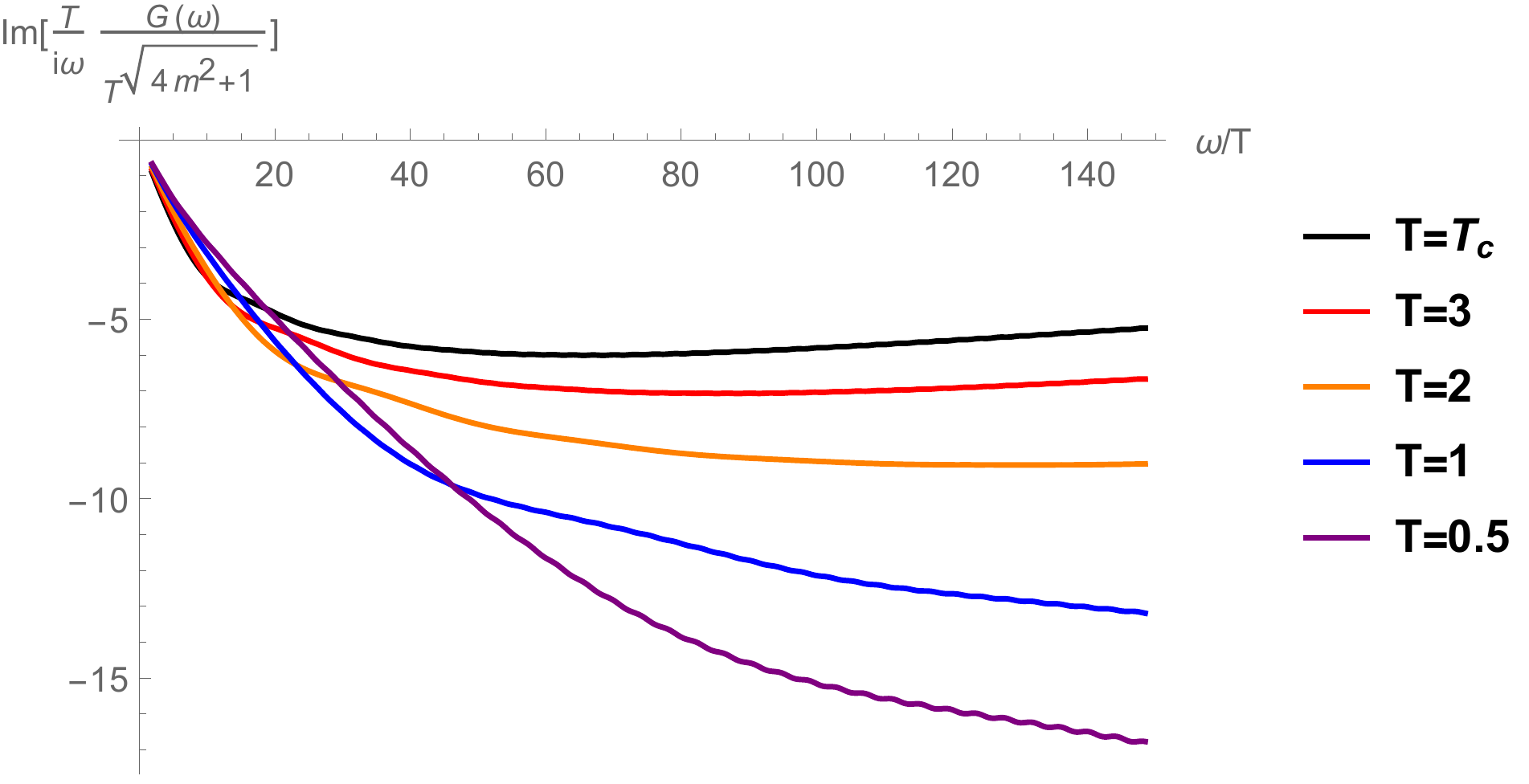}  }   
    \caption{ These figures show the two-point functions with $m^2=0.4$. The curve fitting error of (c) and (d) is given by $\mathbb{E}<1.5\times 10^{-1}$.
} \label{fig:Positive_m_04}
\end{centering}
\end{figure}

\begin{figure}[t] 
	\begin{centering}
		\subfigure[ ]
		{\includegraphics[width=7cm]{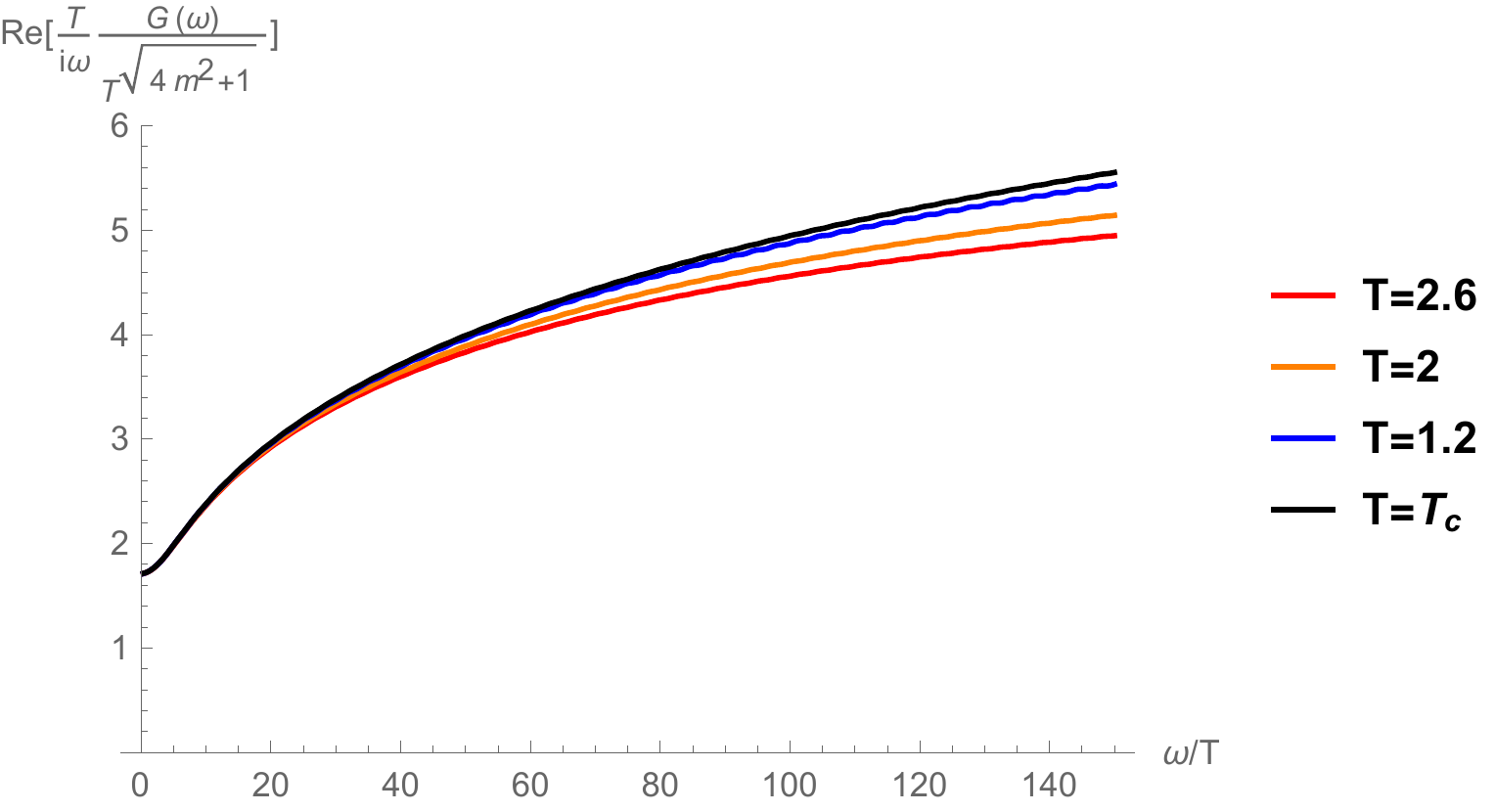}  }
		\subfigure[ ]
		{\includegraphics[width=7cm]{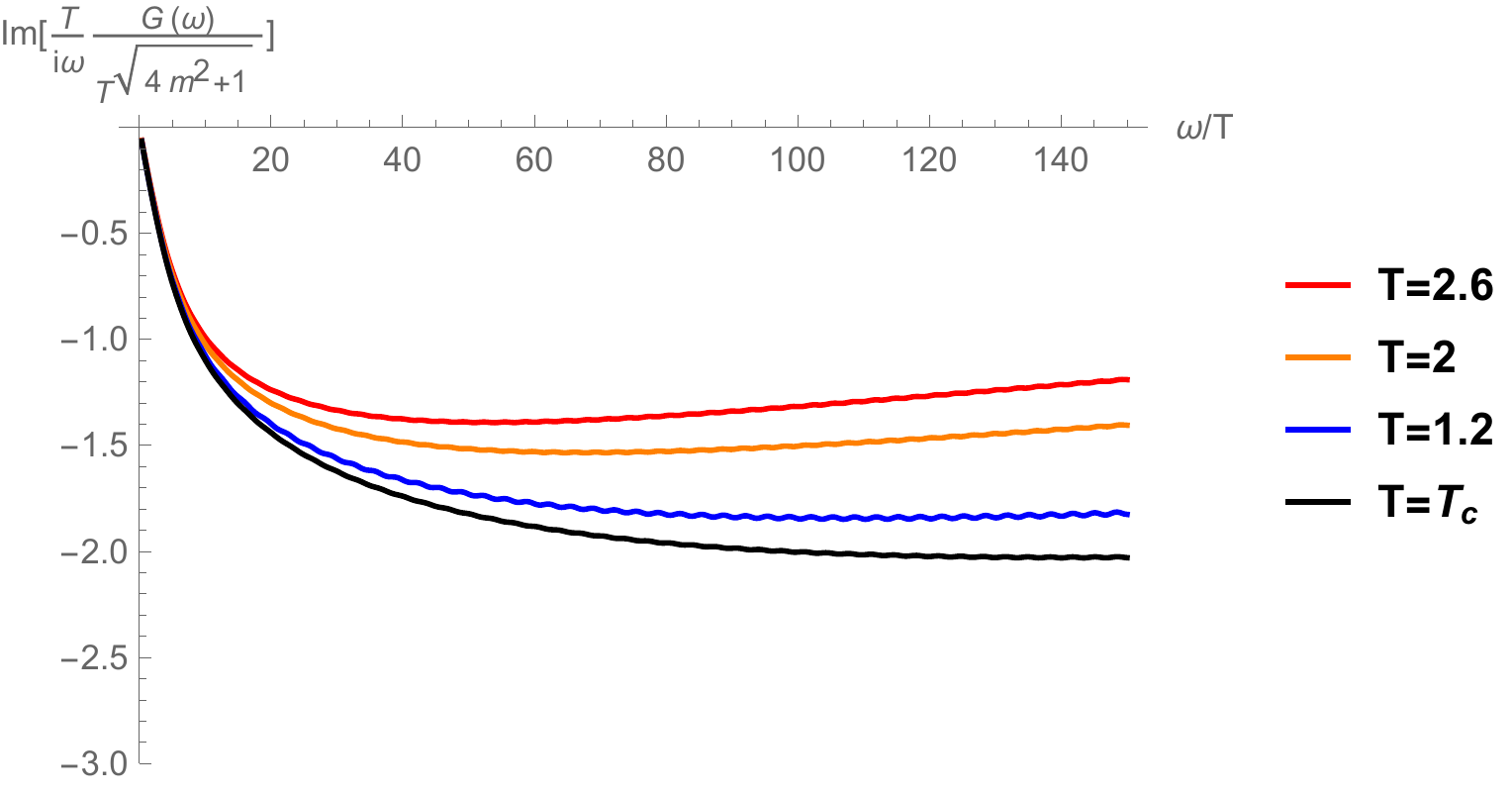}  }\\
		\subfigure[ ]
		{\includegraphics[width=7cm]{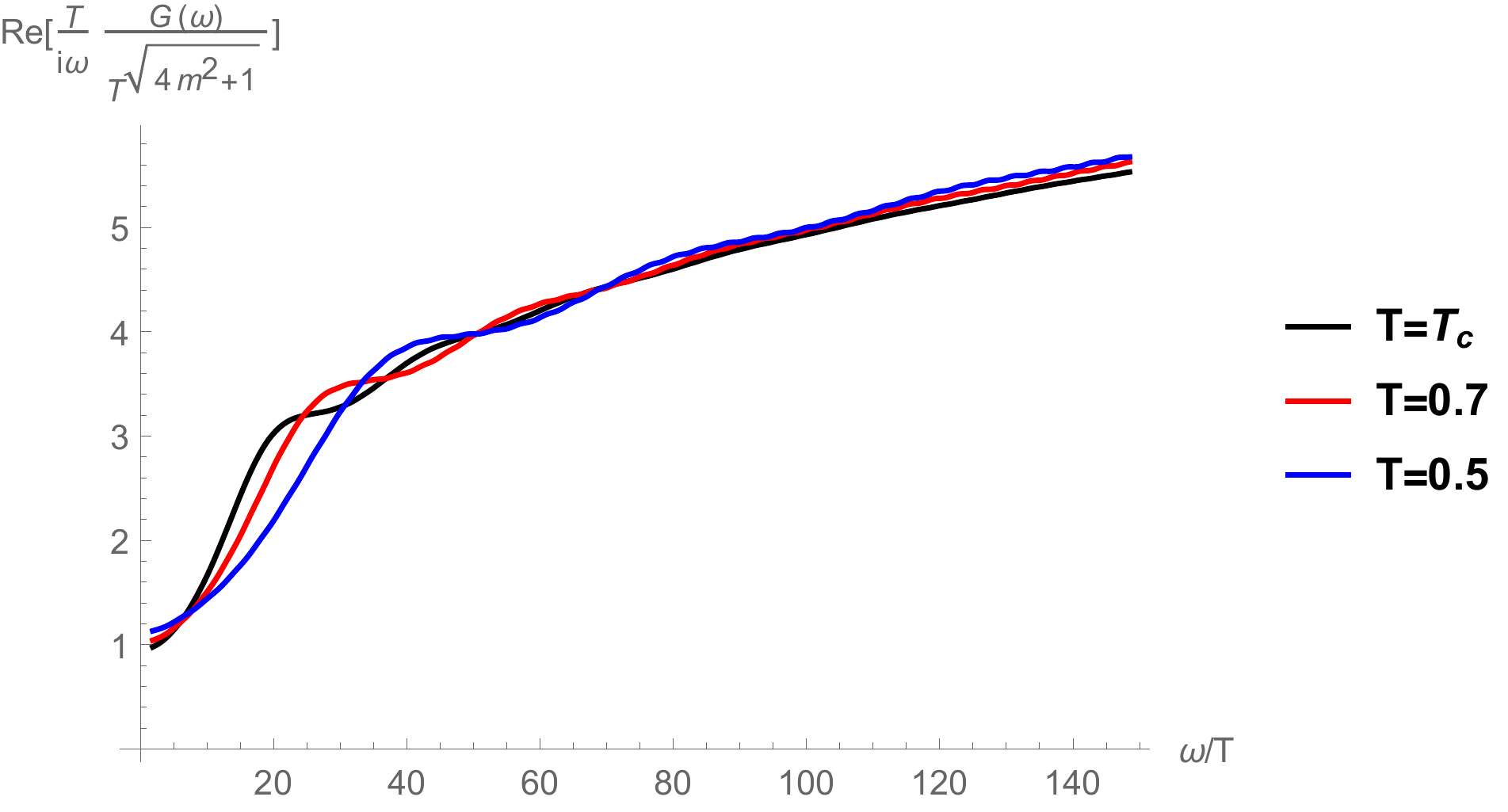}  }
		\subfigure[ ]
		{\includegraphics[width=7cm]{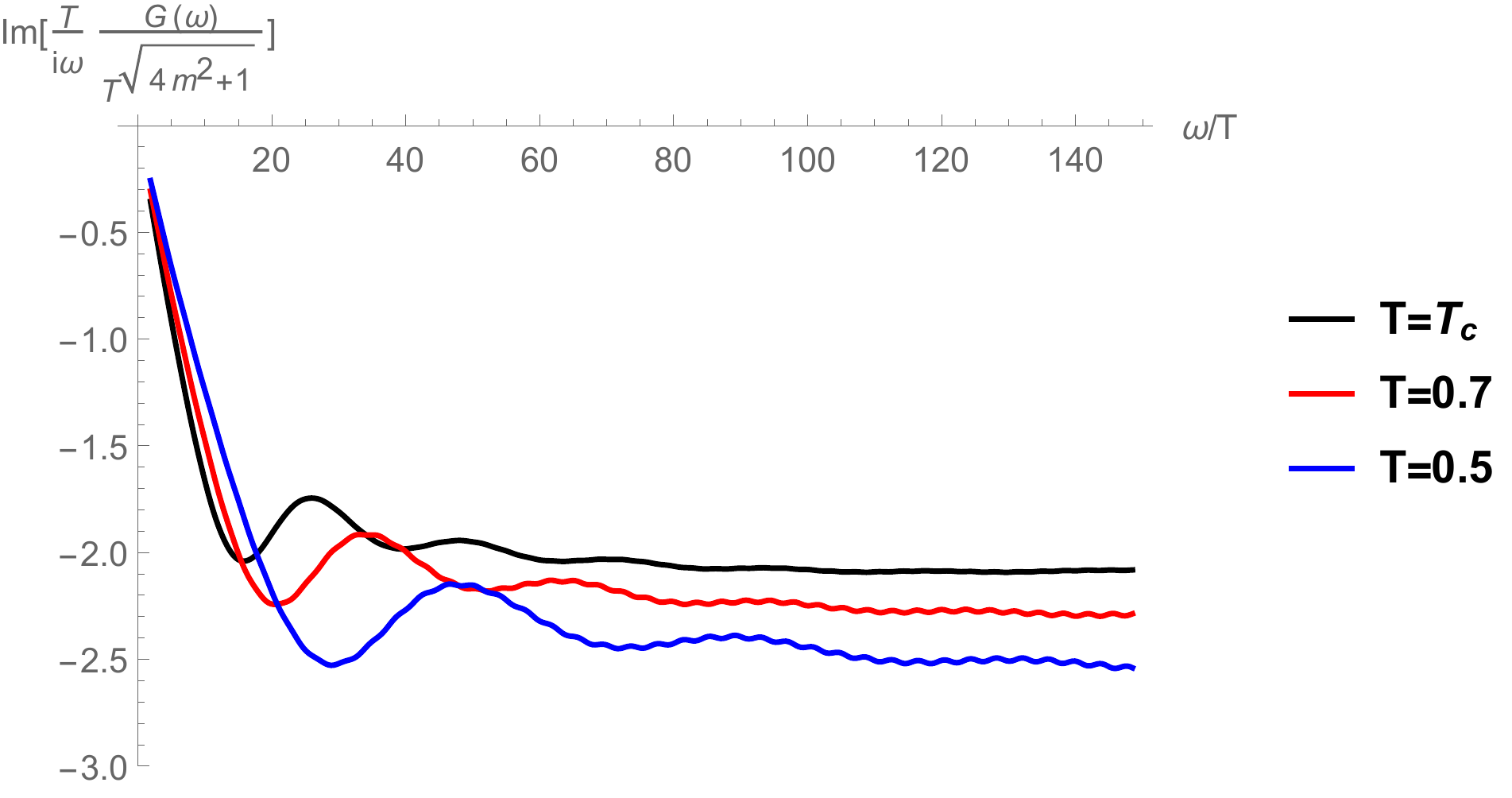}  }   
		\caption{The figures show the two-point function with $m^2=0.2$. The curve fitting error of (c) and (d) is  $\mathbb{E}<6\times 10^{-2}$.
		} \label{fig:Negative_m_02}
	\end{centering}
\end{figure}

\begin{figure}[ht!] 
	\begin{centering}
		\subfigure[ ]
		{\includegraphics[width=7cm]{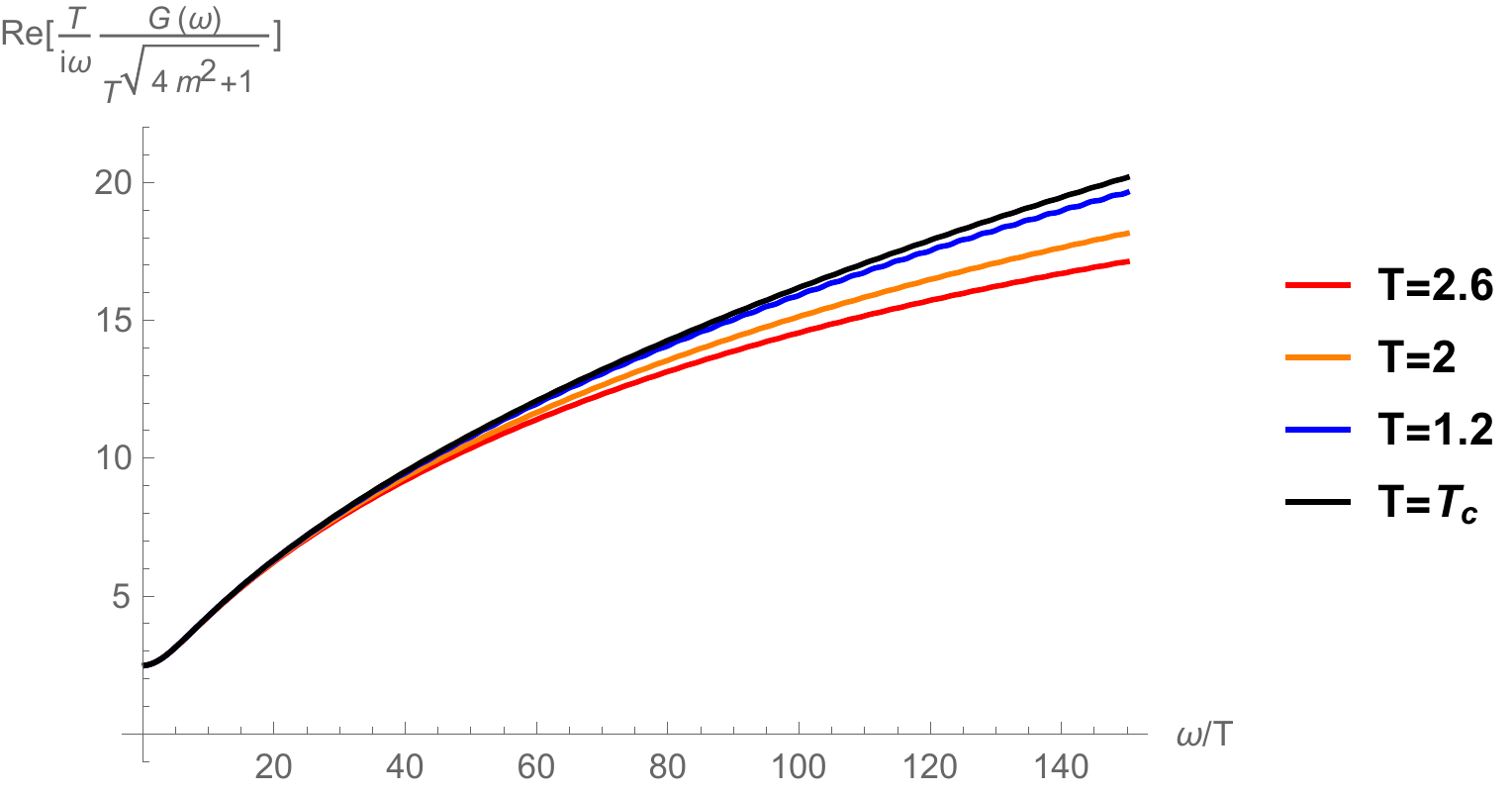}  }
		\subfigure[ ]
		{\includegraphics[width=7cm]{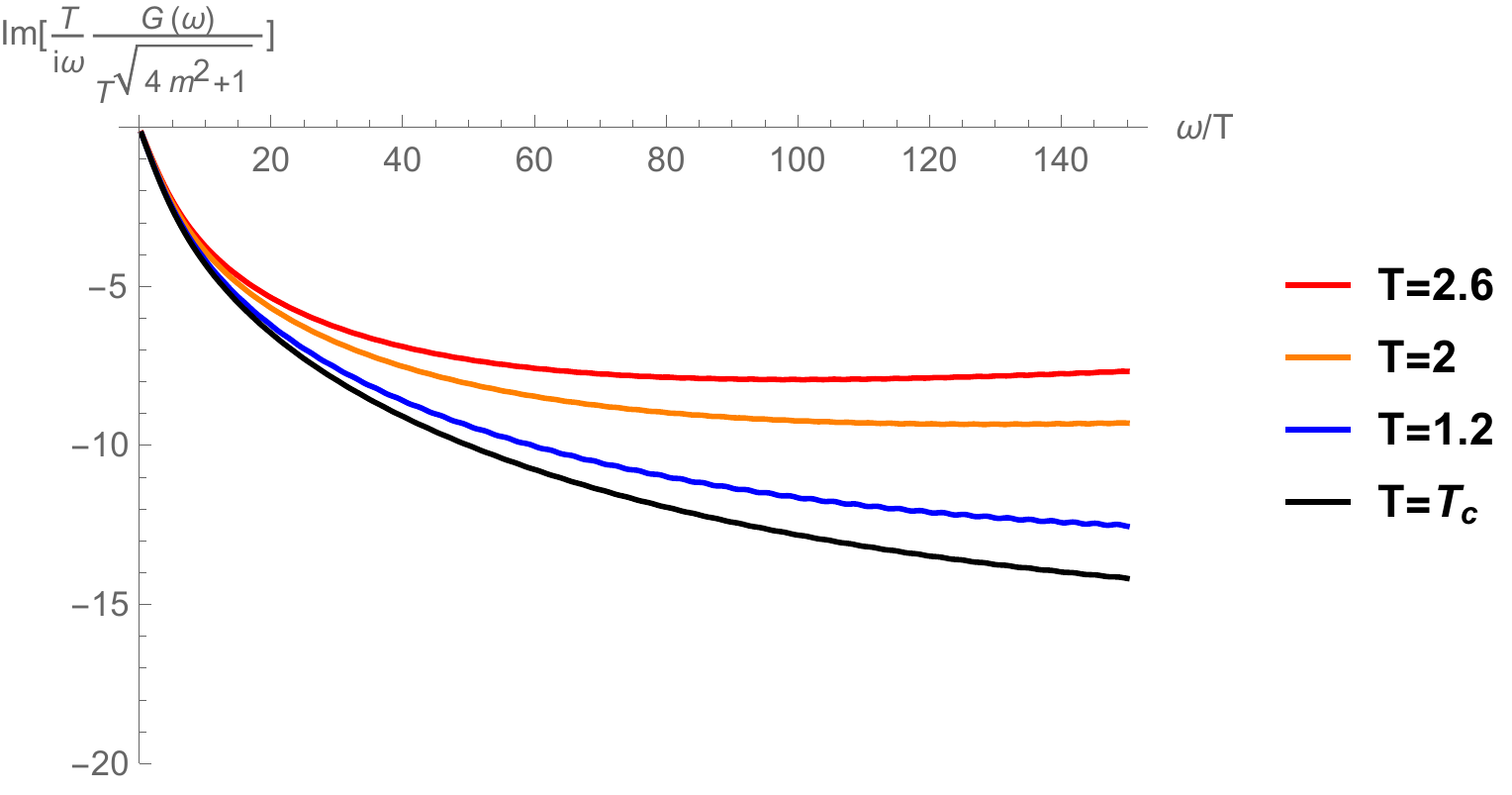}  }\\
		\subfigure[ ]
		{\includegraphics[width=7cm]{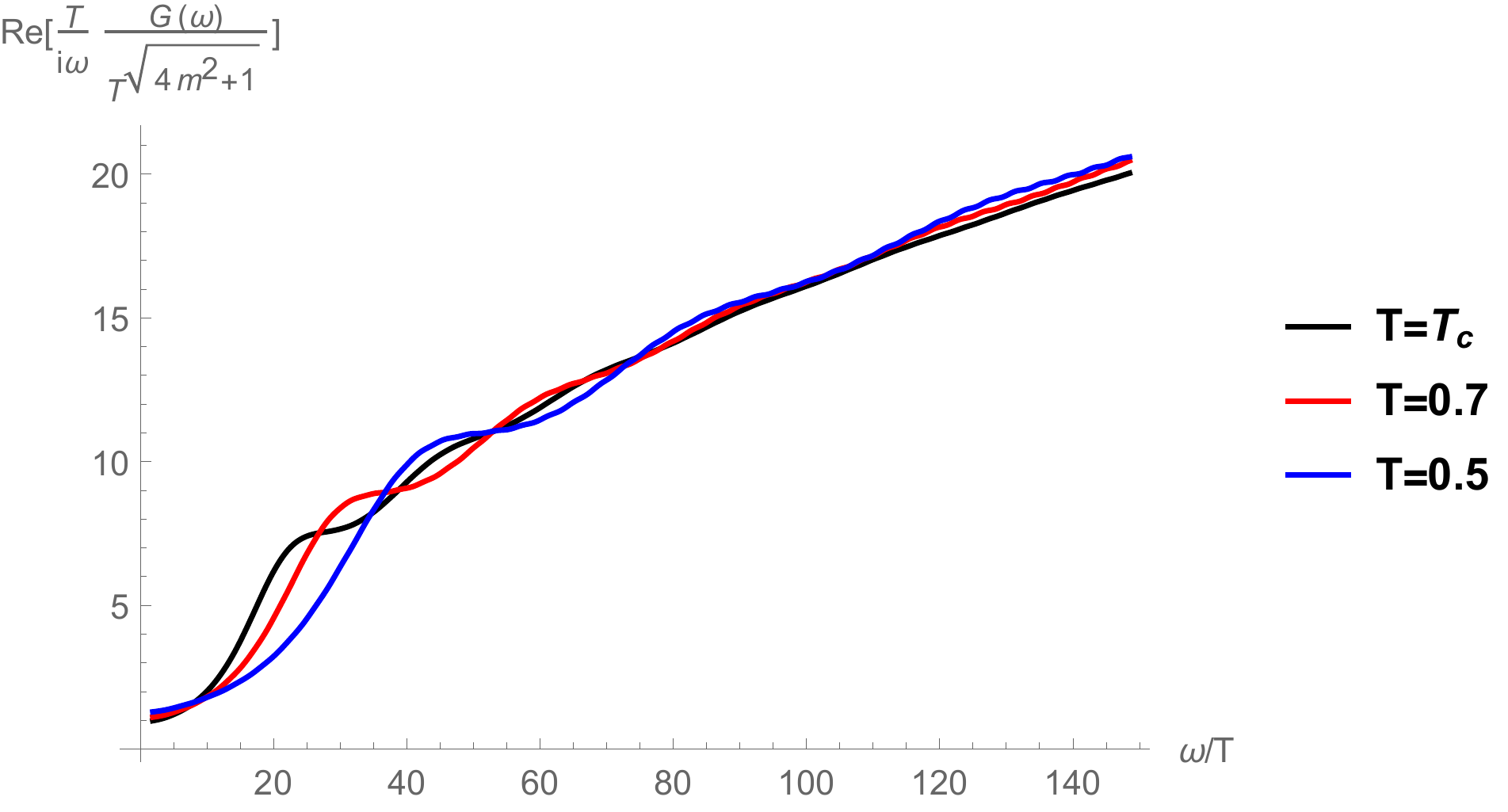}  }
		\subfigure[ ]
		{\includegraphics[width=7cm]{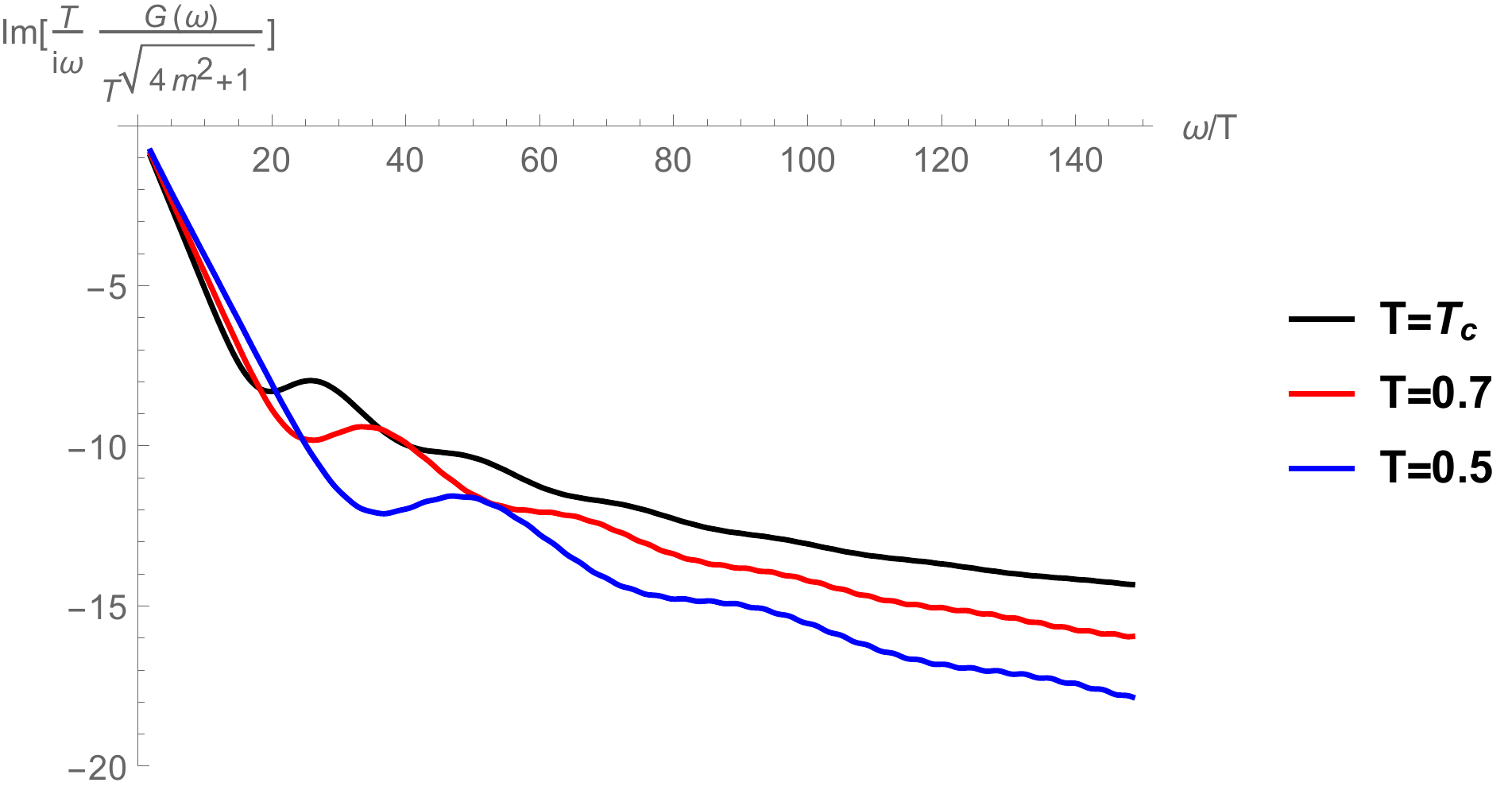}  }   
		\caption{ These figures depict the two-point function, when $m^2=0.4$. The curve fitting error of (c) and (d) is $\mathbb{E}<2.5\times 10^{-1}$.
		} \label{fig:Negative_m_04} 
	\end{centering}
\end{figure}

Together with this boundary condition and the diltaon-potentials (\ref{potential01}) and (\ref{Wnegative}), one can solve (\ref{scalarEq02}) numerically and obtain the relation between source and expectation value. Then, the two-point function can be read off using the following linear response relation 
\begin{align}
G (\omega)\equiv\left< \mathcal{O}(\omega)\mathcal{O}(-\omega) \right> = \frac{\mathcal{O}(\omega)}{J(\omega)}~.
\end{align}
In massive scalar case, a contact term is present in the two-point function. We subtract it and present the numerical results of the two-point functions\footnote{More explicitly, the subtracted Green's function is defined by $\tilde{G}(\omega) = G(\omega)-G(0)$. We drop $``\sim"$ on the subtracted Green function for convenience.} in Figure \ref{fig:Positive_m_02}-\ref{fig:Negative_m_04}.

Before discussing the numerical results, we introduce another point of view to analyze the structure of the two-point function. Using the following tortoise coordinate
\begin{align}
z = {\bar z}{T} \equiv T \int_\infty^r \frac{1}{A(r')}dr'~,
\end{align}
one can rewrite (\ref{scalarEq02}) as 
\begin{align}\label{Schrodinger}
\left(-\frac{d^2}{dz^2}  + V_{\text{eff}}(z)\right)\varphi(z) = \frac{\omega^2}{T^2} \varphi(z)~,  
\end{align}
where the effective potential is given by
\begin{align}\label{Veff}
V_{\text{eff}}(z)= \frac{m^2}{T^2} A(z)~. 
\end{align}
Then, the scalar field equation can be regarded as a Schr$\ddot{\text{o}}$dinger equation with the potential $\frac{m^2}{T^2}A(z)$. The coordinate $z$ ranges from $-\infty$ to $0$. The horizon is located at $z=-\infty$ and $A(-\infty)$ vanishes. Thus, the in-going boundary condition (\ref{in-going}) is equivalent to the left-moving boundary condition at $z=-\infty$:
\begin{align}
\varphi \sim e^{-i \frac{\omega}{T} z}~~~\text{for}~~~ z\to -\infty~.
\end{align}

Now, let us consider the massless case, i.e. $\Delta=1$ case. The potential in (\ref{Schrodinger}) vanishes and the solution is given by the left-moving plane wave solution, $e^{-i\frac{\omega}{T}w}$. The tortoise coordinate can be expanded in terms of original radial coordinate $r$ near the boundary as 
\begin{align}
z\sim T \left( -\frac{1}{r} + \mathcal{O}\left(\frac{1}{r^2}\right) \right)~.
\end{align}
Using this behavior, one can read off the two-point function,
\begin{align}
G^{\Delta = 1}(\omega) = i \omega~.
\end{align}
This is same as the two-point function in $AdS_2$ case. Thus, the temperature doesn't have much effect on the time-time correlation function of the $\Delta=1$ operator, except for the overall scale.

On the other hand, the massive scalar case has much more interesting structure due to the potential $V_{\text{eff}}(z)$. The effect of $V_{\text{eff}}(z)$ on the Schr$\ddot{\text{o}}$dinger equation (\ref{Schrodinger}) becomes smaller in the limit of small mass or high temperature. First interesting case is the situation with  $m^2$ negative but larger than the BF bound. Since the effective potential becomes negative, one may consider negative $\omega^2$ given by a pure imaginary frequency. This implies that an instability could occur below a certain temperature. This instability develops a hairy configuration and generates a real scalar condensation dual to the scalar $\psi$. It would be interesting to study the physical meaning of this phenomenon. Here we focus on the case of positive $m^2$.   

\begin{figure}[t] 
\begin{centering}
    \subfigure[ ]
    {\includegraphics[width=6.5cm]{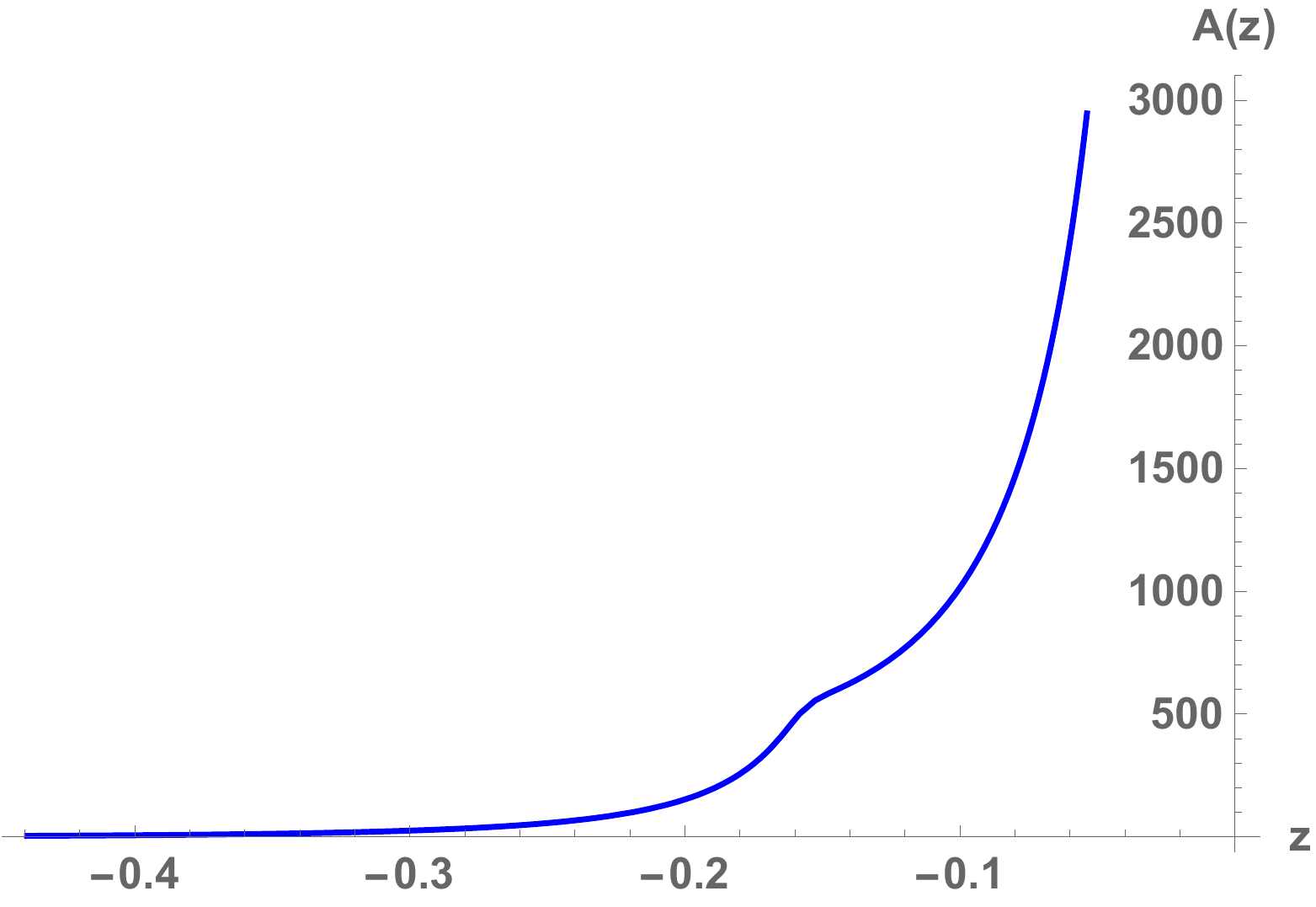}  }
       \subfigure[ ]
   {\includegraphics[width=6.5cm]{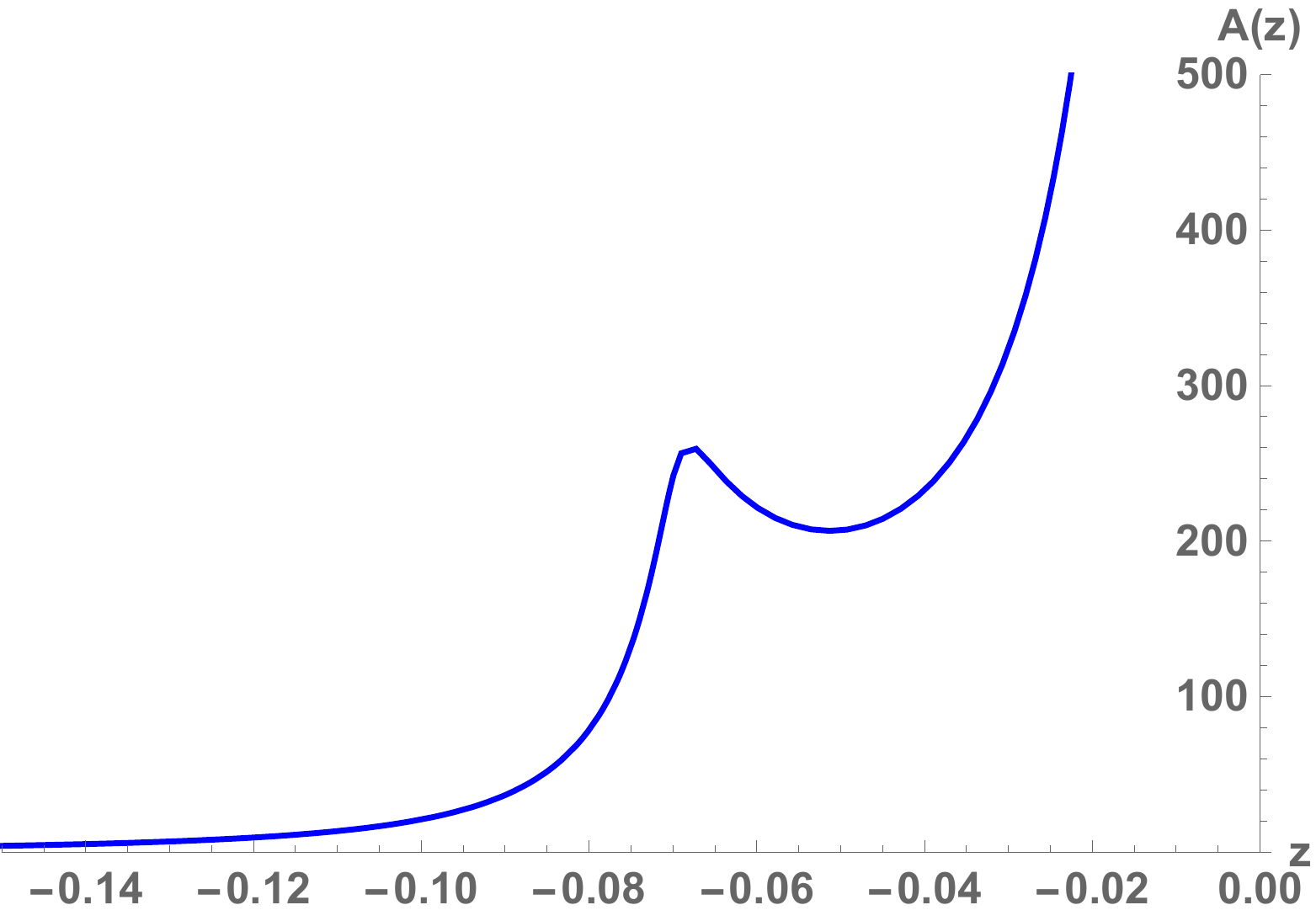}  }  
    \caption{ (a) shows $A(z)$ of the positive potential case at $T=3$ and (b) shows  $A(z)$ of the potential with a negative region at $T\sim 0.47$.
} \label{fig:Az} 
\end{centering}
\end{figure}

A dual operator to a bulk scalar with positive $m^2$ has a dimension $\Delta$ greater than 1. There are two qualitatively different cases. The first is given by the positive diltaon-potential (\ref{potential01}) and the second is the potential with a locally negative region (\ref{Wnegative}). The effective potential $V_{\text{eff}}(z)$ (\ref{Veff}) of the Schr$\ddot{\text{o}}$dinger equation (\ref{Schrodinger}) is proportional to the metric function $A(z)$. As we have shown in Figure \ref{fig:Negative_W}, the locally negative potential gives rise to a local minimum in the metric function below the critical temperatures. This difference also appears in the $V_{\text{eff}}(z)$ of the Schr$\ddot{\text{o}}$dinger equation (\ref{Schrodinger}). See Figure \ref{fig:Az} for some examples. In this figure, $A(z)$ of the locally negative potential yields a local minimum and a potential barrier in $V_{\text{eff}}(z)$. Thus, one can expect the appearance of metastable modes similar to resonances.

\begin{figure}[t] 
\begin{centering}
    \subfigure[ ]
    {\includegraphics[width=7cm]{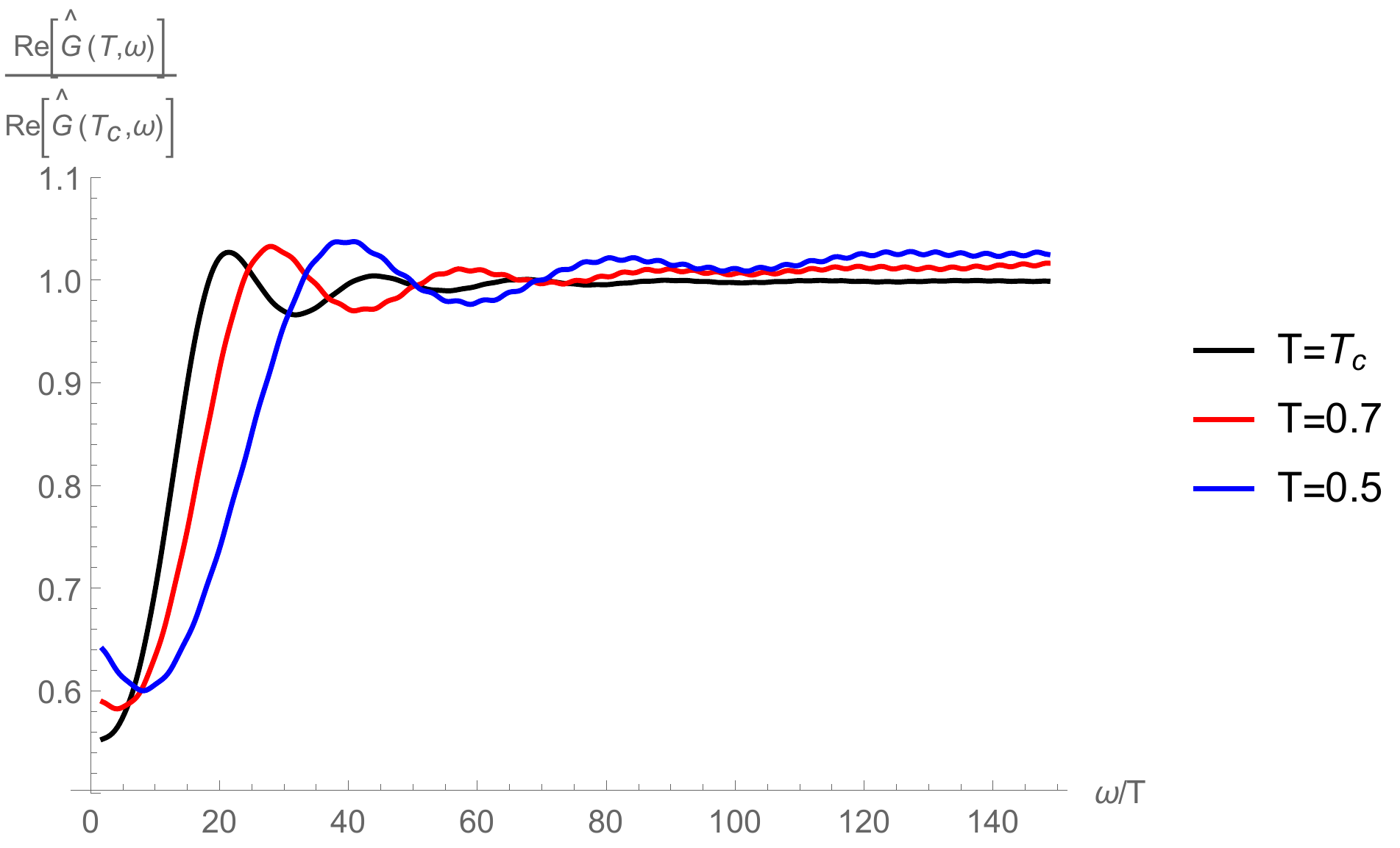}  }
       \subfigure[ ]
   {\includegraphics[width=7cm]{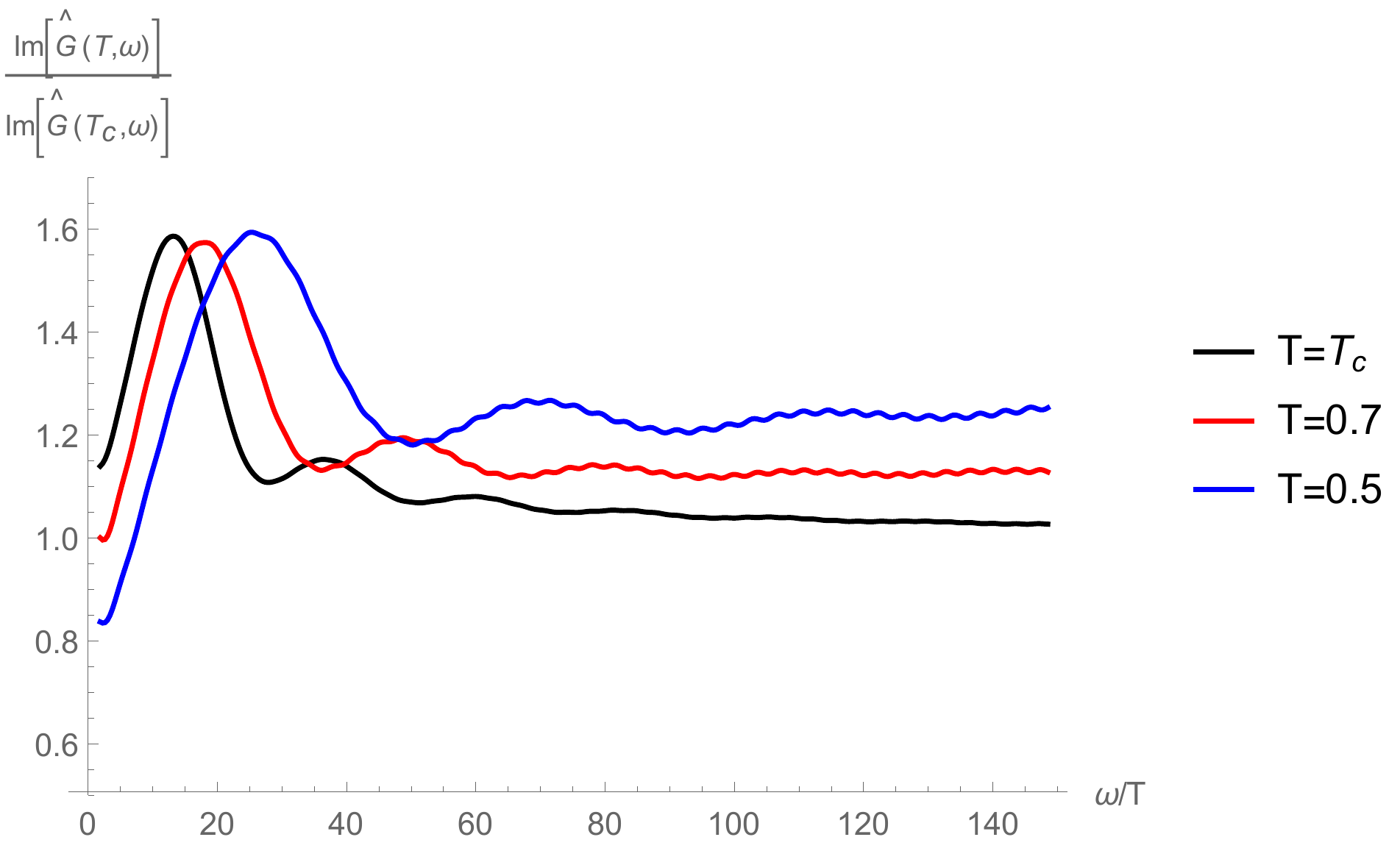}  }\\
     \subfigure[ ]
    {\includegraphics[width=7cm]{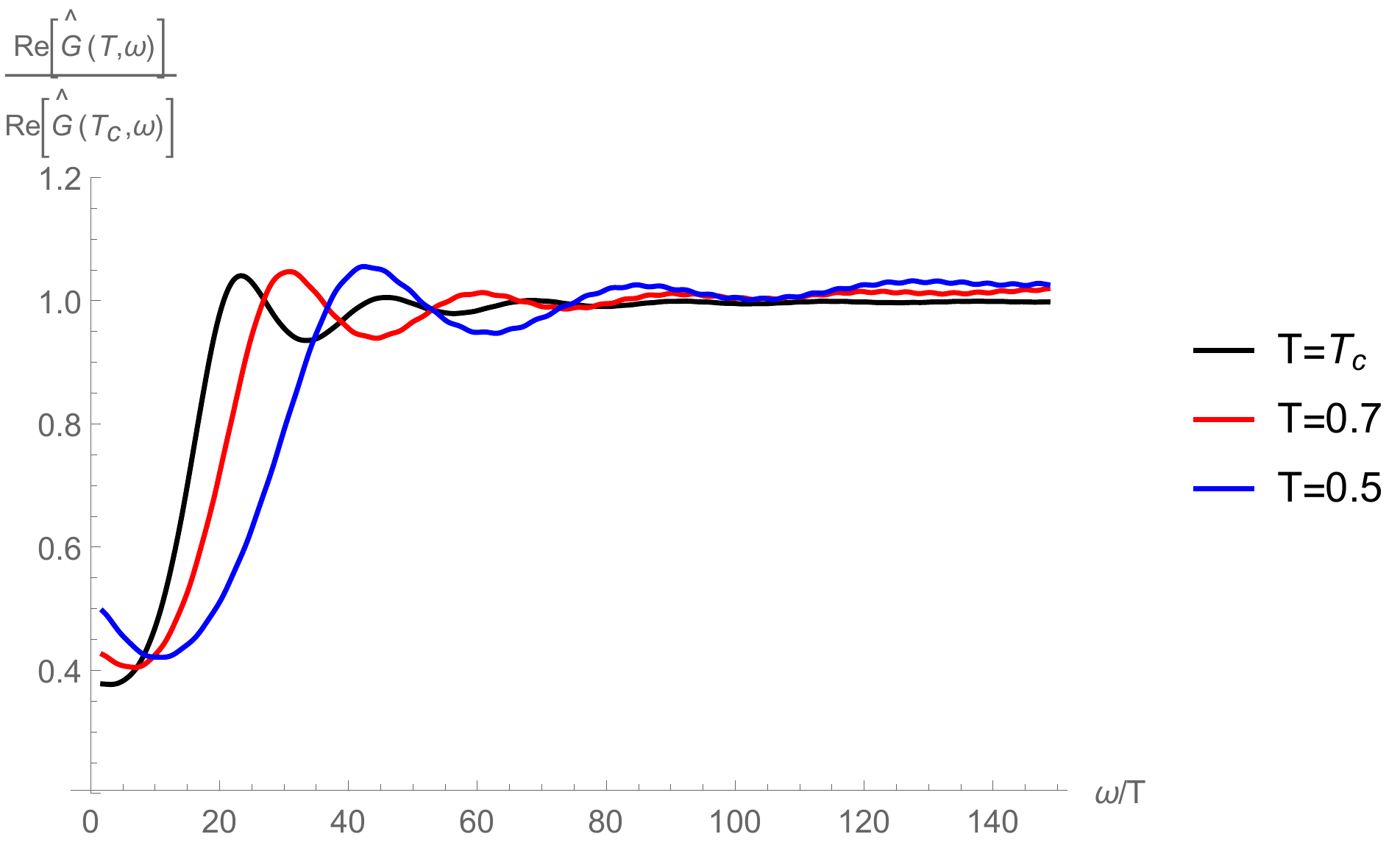}  }
       \subfigure[ ]
   {\includegraphics[width=7cm]{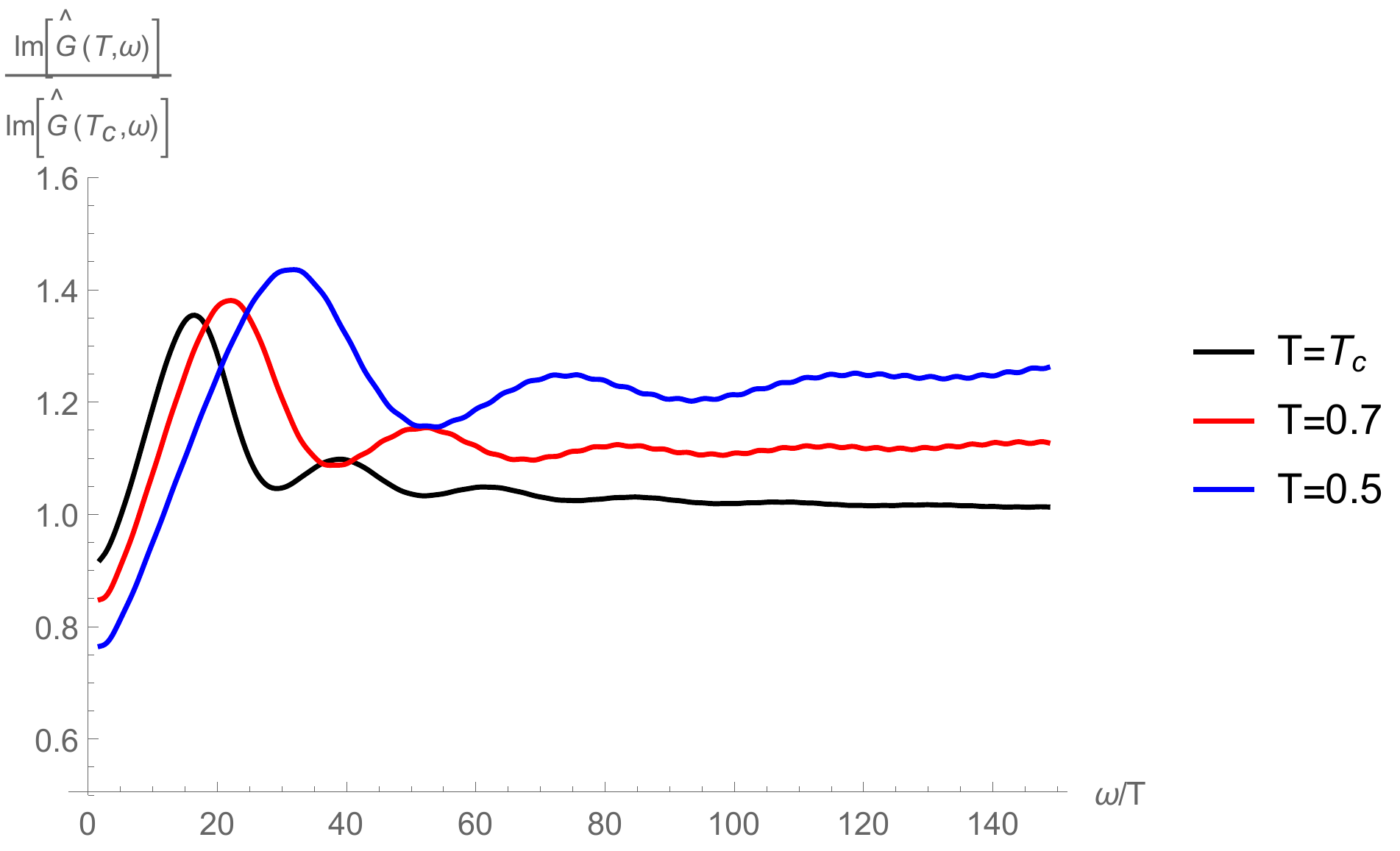}  }   
    \caption{ Here $\hat{G}(T,\omega)$ is defined by $\hat{G}(T,\omega)\equiv \frac{T}{i\omega}\frac{G(\omega)}{T^{\sqrt{4 m^2+1}}}$. All figures show results for the locally negative potential. (a) and (b) correspond to $m^2=0.2$ and (c) and (d) to $m^2=0.4$. The curve fitting errors are as follows.  (a)  $ \mathbb{E} < 1.5\times 10^{-2}$, (b) $\mathbb{E}< 3\times 10^{-2}$,  (c) $\mathbb{E}< 1.5\times 10^{-2}$, (d) $\mathbb{E}< 1.5\times 10^{-2}$.
} \label{fig:ratio}
\end{centering}
\end{figure}

Now, let us look at the effect of locally negative region in the potential at low temperature. The two-point functions below critical temperatures are displayed in (c) and (d) of Figure \ref{fig:Positive_m_02}-\ref{fig:Negative_m_04}. From these figures, one can clearly notice that (c) and (d) of Figure \ref{fig:Negative_m_02} and \ref{fig:Negative_m_04} show peaks in the two-point functions. In order to see these excitations more clearly, we provide ratios of the two-point functions to those just above the critical temperatures in Figure \ref{fig:ratio}. The first peaks are located at $\omega/T = 10\sim 35$. A rough estimation of the energy eigenvalue of the Schr$\ddot{\text{o}}$dinger equation (\ref{Schrodinger}) for these excitations can be calculated by 
\begin{align}\label{estimation00}
\frac{\omega^2}{T^2}\sim \frac{m^2}{T^2} A(x) \sim \frac{m^2}{T^2} \,(250)~,
\end{align}
where we take $250$ for the value of metric function around the local minimum. See Figure \ref{fig:Az} (b). When we consider the masses and temperatures taken in Figure \ref{fig:ratio}, the excitation frequencies $\omega/T$ are estimated around $7$ to $20$. This frequency scale is similar to the locations of the first peaks. It suggests that this resonance-like behavior is originated from the locally negative potential.

Now, let us discuss how this physics is related to the geodesics in the bulk. The scale governing this physics can be read off from (\ref{estimation00}). Except for details of $m^2$, the characteristic scale is roughly given by $\sqrt{A(x_{min})}/T$. This scale has already been introduced as a closed geodesic length in (\ref{scale00}). Therefore, the closed geodesic length of the geometry appears as a scale of the metastable excitations in the two-point function. In addition, as one can see in Figure \ref{fig:ratio}, the peaks of excitations are accompanied by pseudo gap-like behavior. So, all of this physics are related to the scale introduced in (\ref{scale00}). This observation is one of main results in this work.

As a final comment of this section, we point out that for deformed JT gravity, the boundary theory is not known. But at least for a class of deformations, it has been shown that the deformed JT gravity is dual to a matrix
model similar to ordinary JT gravity. In this correspondence, the boundary theory is the
Schwarzian but there are additional ingredients in the bulk due to the deformation \cite{Maxfield:2020ale,Witten:2020wvy}. The
phase transition is expected to appear as a discontinuity of the density of
eigenvalues in a two-cut matrix model. We suppose that our deformed JT gravity model
would be described in similar manner by some matrix model.

%%%%%%%%%%%%%%%
\section{Charged Black Hole}\label{sec3}
%%%%%%%%%%%%%%%

In this section, we discuss a generalization of the effective potential by introducing the electric charge of the black hole \cite{Strominger:1998yg,Hartman:2008dq,Castro:2008ms,Cvetic:2016eiv, Lala:2019inz,Lala:2020lge}. We start with the following action which is an extension of (\ref{Sb}) by a $U(1)$ gauge field $B$ and its field strength $G_{\mu\nu}$,
\begin{align}
S_{Q} = S_{GJT}+  \int_{\mathcal{M}} d^2 x \sqrt{g}\,\frac{1}{4}Z(\phi) G_{\mu\nu}G^{\mu\nu}~.
\end{align}
We assume that $Z(\phi) \geq \phi^{1+\delta}$ for large $\phi$, where $\delta$ is small and positive. In order to find a black hole solution, we take the following ansatz:
\begin{align}
ds^2 = A(x) d\tau^2  + \frac{dx^2}{A(x)}~,~\phi=\phi(x)~,~B = B_\tau (x)\,d\tau~.
\end{align}

The equation of motion for the gauge field can easily be solved as follows:
\begin{align}
B_\tau'(x) = -\frac{Q}{Z(\phi)}~,
\end{align}
where $Q$ is an integration constant. The physical meaning is the charge of the dual quantum mechanics given by 
\begin{align}
Q = - \frac{\delta }{\delta \mu}S_Q^{on-shell} = - \sqrt{g}Z(\phi) G^{x\tau}~,
\end{align}
where $\mu$ is the chemical potential which will be specified below. Plugging this expression into the other equations of motion, the dilaton satisfies $\phi(x)''=0$ like the previous case without gauge field. So we fix $\phi(x)=x$ again. Then, the only equation we have to solve for geometry is 
\begin{align}
A'(x) = W(x) + \frac{Q^2}{Z(x)}~.
\end{align} 
Imposing regularity of the gauge field, we arrive at the charged black hole solution
\begin{align}\label{BtA}
B_\tau(x) = - \int_{x_h}^x dx' \frac{Q}{Z(x')}~~,~~~~~~~A(x) = \int_{x_h}^x dx'\left( W(x') + \frac{Q^2}{Z(x')} \right)~.
\end{align}
So the chemical potential is naturally defined by $\mu = B_\tau (\infty)$. Now, we are ready to discuss thermodynamics of the black hole.

The temperature of the black hole is 
\begin{align}
T = \frac{A'(x_h)}{4\pi} = \frac{1}{4\pi}\left(  W(x_h) + \frac{Q^2}{Z(x_h)} \right).
\end{align} 
Since we are considering asymptotically JT gravity, the asymptotic behavior of the metric function is again given by
\begin{align}
A(x) = x^2 - b + \cdots,
\end{align}
where $b$ is a constant related to the energy. Using (\ref{BtA}), one can derive the following relation for small variation of the parameters:
\begin{align}\label{1stLaw00}
-db = - 4\pi\,d x_h\, T - 2 \mu\, dQ~.
\end{align}
The on-shell action leads to
\begin{align}\label{S_on-shell_00}
S_Q^{on-shell}=& \lim_{\Lambda\to\infty} \frac{\beta}{2}\left\{  \int_{x_h}^{\Lambda} dx \left(\phi_0 A'' + \left(r A' - 2A \right)' + 2\frac{Q^2}{Z(x)} \right)\right.\nonumber\\& \left. ~~~~~~~~~~~~~-   \sqrt{A(\Lambda)}\left(\frac{A'(\Lambda)}{\sqrt{A(\Lambda)}}-2 \right)-\phi_0 A'(\Lambda) \right\}\nonumber\\
=& \beta \left( \frac{b}{2} - ( S_0 + S )T - \mu Q \right)~,
\end{align} 
where $S_0$ is the entropy contribution from $\phi_0$. This on-shell action can be identified with the grand potential $\Omega$ by gauge/gravity correspondence. Comparing (\ref{1stLaw00}) and (\ref{S_on-shell_00}) with the first law and the standard form of the grand potential, one can identify the energy and entropy of the system as 
\begin{align}
E = \frac{b}{2}~~,~~~~~~~S=2\pi x_h\,.
\end{align}  
All of the above identifications lead to the following thermodynamics of the black hole:
\begin{align}
\Omega = T S_Q^{on-shell} = E - (S_0 +S)T -\mu\, Q~~,~~~~~~~dE = TdS + \mu\, dQ\,. 
\end{align}

Now, we discuss the free energy for fixed charge systems. In order to find the difference of free energies as in \cite{Witten:2020ert}, we define the energy of the system as follows:
\begin{align}\label{EnergyExp00}
E(\phi_h)=\lim_{\Lambda\to\infty} \frac{1}{2}\left\{ \Lambda^2 - A(\Lambda) \right\}=\lim_{\Lambda\to\infty}\frac{1}{2}\left\{\Lambda^2-\int_{\phi_h}^{\Lambda}d\phi\left(W(\phi)+ \frac{Q^2}{Z(\phi)}\right) \right\}~.
\end{align}  
From this, one can notice that the energy of the dual system is described by the effective potential:
\begin{align}
W_Q(\phi) \equiv W(\phi)+ \frac{Q^2}{Z(\phi)}~.
\end{align}
Eq.(\ref{EnergyExp00}) implies that any black hole geometry with $W'(x_h)<0$ has negative a heat capacity. Thus, the physically relevant black hole geometries satisfy $W'(x_h)>0$. On the other hand the free energy difference in the canonical ensemble for  fixed charge systems is given by
\begin{align}
\Delta F = \Delta E - T \Delta S =2\pi  \int_{\phi_L}^{\phi_R} d\phi (T_Q(\phi)-T)~,
\end{align}
where $T_Q (\phi)= \frac{1}{4\pi}W_Q(\phi)$, and $\phi_L$ and $\phi_R$ denote two possible locations of the horizon with $\phi_L < \phi_R$. The arguments for the phase transition and the two-point function in section \ref{sec2} can be applied to this charged black hole. The only difference is that we have to consider $W_Q(\phi)$ instead of $W(\phi)$.

%%%%%%%%%%
\section{Local $T\bar{T}$ Deformation from 2D Gravity}\label{sec4}
%%%%%%%%%%%

In this section, we introduce a different perspective of the generalized JT gravity using a $T\bar{T}$ deformation. This is an extension of the constant $W(\phi)$ case which was studied in \cite{Dubovsky:2018bmo}.

Let us start with the generalized JT gravity (\ref{Sb}) coupled to conformal matter for which the action is given by 
\begin{align}
S_{T\bar{T}}=S_{GJT} + S_m(e_\mu^a,\Psi)~,
\end{align}
where $e_\mu^a$ is the zweibein and $\Psi$ is the conformal matter field. We use the first order formalism to rewrite the action in terms of the zweibein $e^a_\mu$, spin connection $w_\mu{}^a{}_b=w_\mu\epsilon^a{}_b$, and Lagrange multiplier $\sigma_a$. In the two-dimensional space, the scalar curvature is given by $eR=\epsilon^{\mu\nu}(\p_\mu w_\nu-\p_\nu w_\mu)$, where $e=\text{det}(e^a{}_\mu)$. Using this expression, one can rewrite the gravity action as follows: 
\begin{align}
S_{GJT}=&  -\frac{1}{16\pi G_N}\int d^2x\,  \epsilon^{\mu\nu}\left( 2\phi\p_\mu w_\nu + \frac{W(\phi)}{2}\epsilon_{ab}e^a{}_\mu e^b{}_\nu  -\sigma_a(\p_\mu e^a{}_\nu+ w_\mu\epsilon^a{}_be^b{}_\nu)\right) \,,
\end{align}
where we have dropped the total derivative terms and introduced the Newton constant.

Now, we consider the equations of motion to show this action is related to a deformed conformal field theory. The equation of motion for $\phi$ is given by
\begin{equation}\label{phieomdJT1}
\e^{\mu\nu}\left(2\p_\mu w_\nu +\frac{\p_{\phi}W(\phi)}{2}\epsilon_{ab}e^a{}_\mu e^b{}_\nu\right)=0\,.
\end{equation} 
Therefore, the scalar curvature satisfies 
\begin{equation}\label{SCurvature}
e(R[w]+\p_{\phi}W(\phi))=0 \,.
\end{equation}
From these relations, one can notice that $\omega_\mu$ is a function of $\phi$ and $e$ and the scalar curvature is a function of $\phi$ only. In general, the metric of the two-dimensional space can be written in a conformally flat form. When the geometry has one Killing vector, one can always choose $e$ to be a constant by a coordinate transformation. For simplicity, we take $e=1$ as in the previous black hole geometry. Then, the spin connection $\omega_\mu$ becomes a function of $\phi$ only.

Plugging the equation  \eqref{phieomdJT1} into the action, the full action becomes
\begin{equation}
S_{T\bar{T}} = -\frac{1}{16\pi G_N}\int d^2x\,\epsilon^{\mu\nu}\left(\frac{1}{2}(W-\phi \p_{\phi}W)\epsilon_{ab}e^a{}_\mu e^b{}_\nu + D_\mu\sigma_ae^a{}_\nu \right) + S_m(e_\mu^a,\Psi) \,,
\end{equation}
where $D_\mu\sigma_a=\p_\mu\sigma_a + w_\mu\epsilon_a{}^b\sigma_b$. Let $f=W-\phi\p_{\phi}W$. The action can be written in factorized form as
\begin{align}
S_{T\bar{T}} =&-\frac{1}{16\pi G_N}\int d^2x\,\frac{f}{2}\epsilon^{\mu\nu}\epsilon_{ab}\left(e^a{}_\mu  -\frac{1}{f}\epsilon^{ac}D_\mu\sigma_c\right)\left(e^b{}_\nu - \frac{1}{f}\epsilon^{bd}D_\nu\sigma_d\right) \nonumber\\
&+ \frac{1}{16\pi G_N}\int d^2x\,\frac{1}{2f}\,\epsilon^{\mu\nu}\epsilon_{ab}(\epsilon^{ac}D_\mu\sigma_c)(\epsilon^{bd}D_\nu\sigma_d) +S_m \,.
\end{align} 
 If we rescale $\sigma_a$ to $f\sigma_a$, the covariant derivative changes to
\begin{equation}
D_\mu\sigma_c \rightarrow f\left(D_\mu+ f^{-1}\p_\mu f \right)\sigma_c \equiv f\tilde{D}_\mu\sigma_c \,,
\end{equation}
where we have defined the new covariant derivative $\tilde{D}_\mu$ including the additional term of $\p_\mu\log f$. Note there is no $e^a{}_\mu$ dependence in the new covariant derivative.
Introducing the notation 
\begin{equation}\label{Dcoordinate}
X^a = \e^{ab}\sigma_b ~~,~~~~~ \tilde{e}^a_\mu=\tilde{D}_\mu X^a = \partial_\mu X^a + \mathcal{A}_\mu{}^a{}_c X^c,
\end{equation}
where $ \mathcal{A}_\mu{}^a{}_c\equiv \omega_\mu\epsilon^a{}_c + \delta^a_c \partial_\mu \log f $, 
the action can be written as
\begin{align}\label{Sinter}
S_{T\bar{T}} =&-\frac{1}{16\pi G_N}\int d^2x\,\frac{f}{2}\epsilon^{\mu\nu}\epsilon_{ab}\left(e^a{}_\mu - \tilde{e}^a{}_\mu\right)\left(e^b{}_\nu - \tilde{e}^b{}_\nu\right) \nonumber\\
&+\frac{1}{16\pi G_N}\int d^2x\,\frac{f}{2}\,\epsilon^{\mu\nu}\epsilon_{ab}\tilde{e}^a{}_\mu\tilde{e}^b{}_\nu +S_m \,.
\end{align}

Now, the equation of motion for $X^a$ is given by
\begin{align}
\tilde{D}'_\mu \frac{\delta S_{T\bar{T}}}{\delta \tilde{e}^a{}_\mu}=\partial_\mu \frac{\delta S_{T\bar{T}}}{\delta \tilde{e}^a{}_\mu} - \mathcal{A}_\mu{}^c{}_a \frac{\delta S_{T\bar{T}}}{\delta \tilde{e}^c{}_\mu}= 0
\end{align}
Thus, we can define the energy-momentum tensor as the variation with respect to $\tilde{e}^a{}_\mu$: 
\begin{equation}\label{defTeom}
 \tilde{e}\tilde{T}^\mu{}_a \equiv \frac{\d S_{T\bar T}}{\d \tilde{e}^a{}_\mu} = -\frac{f}{16\pi G_N}\e^{\mu\nu}\e_{ab}e^b{}_\nu.
\end{equation}
The $e^a{}_\mu$ equation of motion gives
\begin{equation}\label{Teom}
\frac{\d S}{\d e^a{}_\mu} = -\frac{f}{16\pi G_N}\e^{\mu\nu}\e_{ab}(e^b{}_\nu - \tilde{e}^b{}_\nu) + eT^\mu{}_a = 0,
\end{equation}
where we have defined the matter stress tensor as
\begin{equation}
\frac{\delta S_m}{\delta e^a{}_\mu} = eT^\mu{}_a,
\end{equation}
so that
\begin{equation}\label{e-e}
e^a{}_\mu - \tilde{e}^a{}_\mu = \frac{16\pi G_N}{f}\e_{\mu\nu}\e^{ab}eT^\nu{}_b.
\end{equation} 
The action \eqref{Sinter} can be written in terms of tilded fields via \eqref{defTeom} 
\begin{eqnarray}\label{Stilde}
S_{T\bar{T}} 
%&=& -\frac{1}{16\pi G_N}\int d^2x\,\frac{f}{2}\e^{\mu\nu}\e_{ab}\left(\frac{16\pi G_N}{f}\e_{\mu\rho}\e^{ac}\,eT^\rho{}_c\right)\left(\frac{16\pi G_N}{f}\e_{\nu\sigma}\e^{bd}\,eT^\sigma{}_d\right) \nn\\
%&&+ \frac{1}{16\pi G_N}\int d^2x\,\frac{f}{2}\,\epsilon^{\mu\nu}\epsilon_{ab}\,\tilde{e}^a{}_\mu\tilde{e}^b{}_\nu +S_m(e_*) \nn\\
%&=& -\frac{1}{16\pi G_N}\int d^2x\,\frac{f}{2}\e^{\mu\nu}\e_{ab}\left(\frac{16\pi G_N}{f}\e_{\mu\rho}\e^{ac}\,\tilde{e}\tilde{T}^\rho{}_c + \tilde{e}^a{}_\mu\right)\left(\frac{16\pi G_N}{f}\e_{\nu\sigma}\e^{bd}\,\tilde{e}\tilde{T}^\sigma{}_d + \tilde{e}^b{}_\nu\right) \nn\\
%&&+ \frac{1}{16\pi G_N}\int d^2x\,\frac{f}{2}\,\epsilon^{\mu\nu}\epsilon_{ab}\,\tilde{e}^a{}_\mu\tilde{e}^b{}_\nu +S_m(e_*) \nn\\
&=& -\int d^2x\,\left(\frac{16\pi G_N}{2f}\e_{\rho\sigma}\e^{cd}\,\tilde{e}^2\tilde{T}^\rho{}_c\tilde{T}^\sigma{}_d + \tilde{e}\tilde{T}^\rho{}_c\tilde{e}^c{}_\rho \right) + S_m
\end{eqnarray}

The trace of the deformed stress tensor in the second term can be computed from the condition that the stress tensor of the matter action is traceless $e^a{}_\mu T^\mu{}_a=0$, which yields through \eqref{Teom}, \eqref{defTeom},
\begin{equation}
0=-\frac{f}{16\pi G_N}\e^{\mu\nu}\e_{ab}\,e^a{}_\mu(e^b{}_\nu - \tilde{e}^b{}_\nu) = \tilde{e}\tilde{T}^\nu{}_b\left( -\frac{16\pi G_N}{f}\e_{\nu\sigma}\e^{bc}\,\tilde{e}\tilde{T}^\sigma{}_c - \tilde{e}^b{}_\nu\right) .
\end{equation}
Thus, we find the trace relation
\begin{equation}
\tilde{T}^\nu{}_b\tilde{e}^b{}_\nu = -\frac{16\pi G_N}{f}\e_{\nu\sigma}\e^{bc}\,\tilde{e}\tilde{T}^\nu{}_b\tilde{T}^\sigma{}_c.
\end{equation}
Then, the action \eqref{Stilde} becomes  
\begin{eqnarray}
S_{T\bar{T}} &=& \int d^2x \frac{16\pi G_N}{2f}\e_{\mu\nu}\e^{ab}\,\tilde{e}^2\tilde{T}^\mu{}_a\tilde{T}^\nu{}_b + S_m(e^a_{*\mu}) \nn\\
&=& \int d^2x \,\tilde{e}\,\l\det(\tilde{T}_{ab}) + S_m(e^a_{*\mu}),
\end{eqnarray}
where $e^a_{*\mu}$ is the on-shell zweibein and we have defined the deformation parameter 
\begin{align}\label{lambdaC}
\lambda = \frac{16\pi G_N}{f}=\frac{16\pi G_N}{W-\phi \partial_\phi W}~.
\end{align}
Variation of the action with respect to $\lambda$ yields
\begin{eqnarray}
\frac{\d S_{T\bar{T}}}{\d \l} &=& \int d^2x\,\frac{\delta S_{T\bar{T}}}{\delta e_*^a{}_{\mu}}\frac{\delta e_*^a{}_{\mu}}{\delta \l} + \tilde{e} \det(\tilde{T}_{ab}) \nn\\
&=& \tilde{e}\det(\tilde{T}_{ab}) \,,
\end{eqnarray}
where the on-shell condition has been used. This can be interpreted as the $T\bar{T}$ deformation. The deformed zweibein is now given by $\tilde{e}_\mu^a$ so the $X^a$ in (\ref{Dcoordinate}) plays a role of a dynamical coordinate discussed in numerous literature \cite{Dubovsky:2017cnj, Cardy:2018sdv, Conti:2018tca,Aguilera-Damia:2019tpe,Tolley:2019nmm,Mazenc:2019cfg,Caputa:2020lpa}.

Note that the deformation parameter is now a function of the dilaton $\phi(x)$. Thus, the deformation parameter is a local function except for a special case. A simple example is given by a potential, $W(\phi)=\Lambda + \Sigma\, \phi$, which was studied in \cite{Almheiri:2014cka}. This potential can describe two gravity theories. One is the JT gravity with positive $\Sigma$ and $\Lambda=0$. The other one is the gravity model with a constant potential, i.e, $\Sigma=0$. The relation to the $T\bar{T}$ deformation of the latter case was clarified in \cite{Dubovsky:2018bmo}. The corresponding deformation parameter is given by a constant $\lambda = \frac{16\pi G_N}{\Lambda}$ and this gravity model describes a flat space. Therefore, this flat JT gravity coupled to a conformal matter is equivalent to the constant coupling $T\bar{T}$ deformation of the conformal theory at the classical level.

On the other hand, the nonvanishing $\Sigma$ case describes a curved space due to (\ref{phieomdJT1}) and (\ref{SCurvature}), although it has the same constant deformation parameter $\lambda$. In fact, this model is equivalent to JT gravity up to a total derivative. This can be seen by a field redefinition, $\phi\to\phi - \frac{\Lambda}{\Sigma}$. Therefore, JT gravity coupled to matter can be interpreted as a $T\bar{T}$ deformed conformal field theory with a constant deformation parameter $\lambda = \frac{16\pi G_N}{\Lambda}$.

\begin{figure}[t] 
\begin{centering}
    \subfigure[ ]
    {\includegraphics[width=4.8cm]{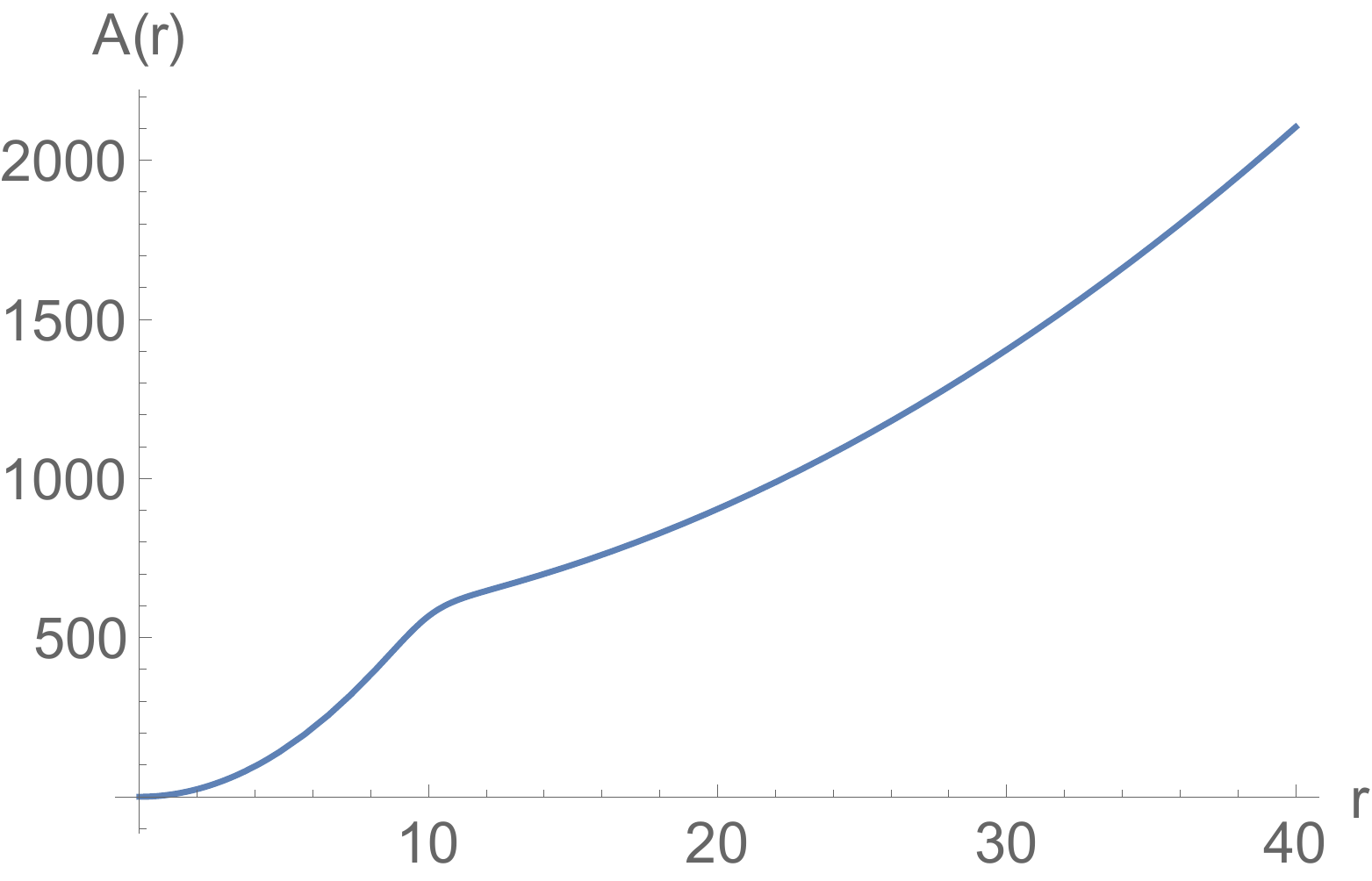}  }
       \subfigure[ ]
   {\includegraphics[width=4.8cm]{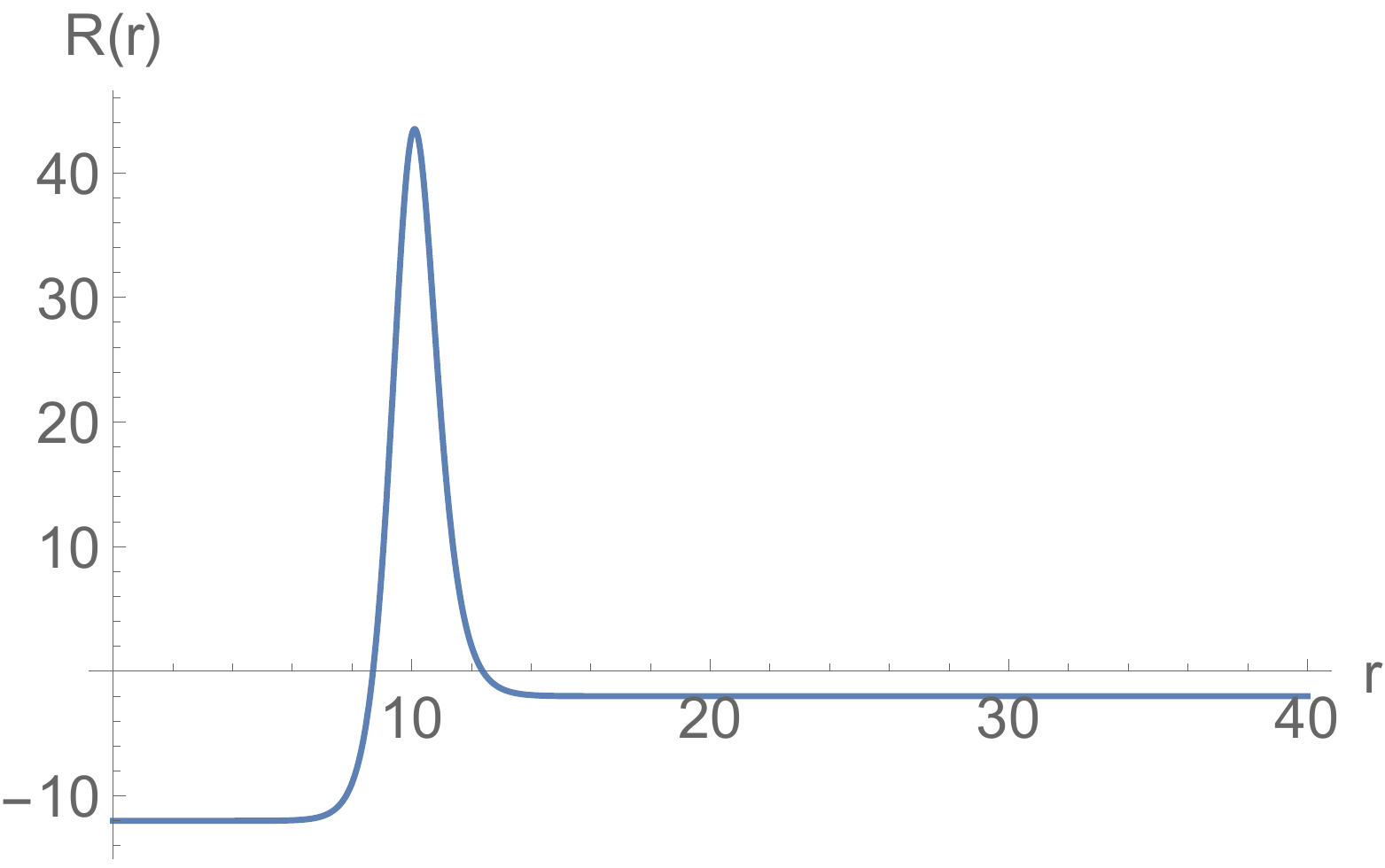}  } 
     \subfigure[ ]
    {\includegraphics[width=4.8cm]{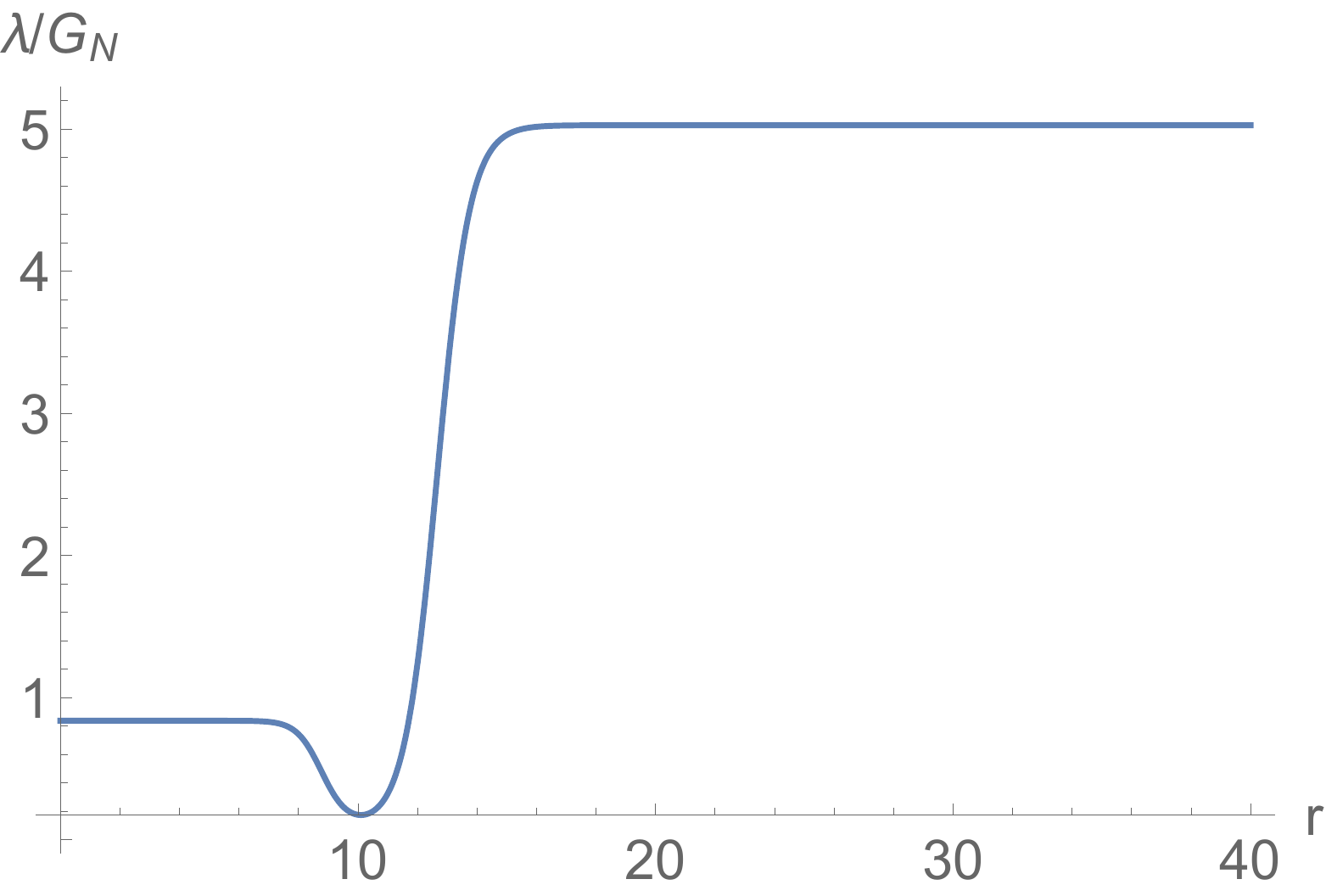}  }
    
    \caption{ Plots of $A(r)$, $R(r)$, and $\l/G_N$ for $s=5$.
} \label{fig:AR_lambda}
\end{centering}
\end{figure}

In general, the deformation coupling can be a local function. To see this more explicitly, we choose a potential introduced in (\ref{potential01}). However, as we discussed, the potential of this model becomes $W(\phi)\sim 2\phi$ for large $\phi$ and $W(\phi)\sim 12\phi$ for small $\phi$. In these regions, the corresponding deformation coupling $\lambda$ in (\ref{lambdaC}) diverges. In order to avoid this divergence, we consider field redefinition $\phi\to\phi-s$ as above and then the horizonless vacuum geometry is given by the integration from $-s$ as follows:
\begin{align}
A(x) = \int_{-s}^x d\phi\, W(\phi + s)~.
\end{align}
In this case, the radial coordinate is nothing but $r=x+s$ with $r\geq 0$. This geometry is a domain wall interpolating different (Euclidean) AdS spaces with different cosmological constants. The explicit form of the metric function is shown in Figure \ref{fig:AR_lambda} (a). The analytic expression is given in (\ref{DomainW}). Also, we display the scalar curvature and corresponding coupling function $\lambda$ in Figure \ref{fig:AR_lambda}. Note that the corresponding $T\bar{T}$ deformation is a Janus type. 

%For the case of a locally negative potential, the coupling constant $\lambda$ has divergences at some points. This can't be avoided by a field redefinition of the dilaton. This issue is not fully understood yet. Since this is just a classical analysis, it would be interesting to clarify this point with quantum corrections.     

\begin{figure}[t] 
	\begin{centering}
		\subfigure[ ]
		{\includegraphics[width=4.8cm]{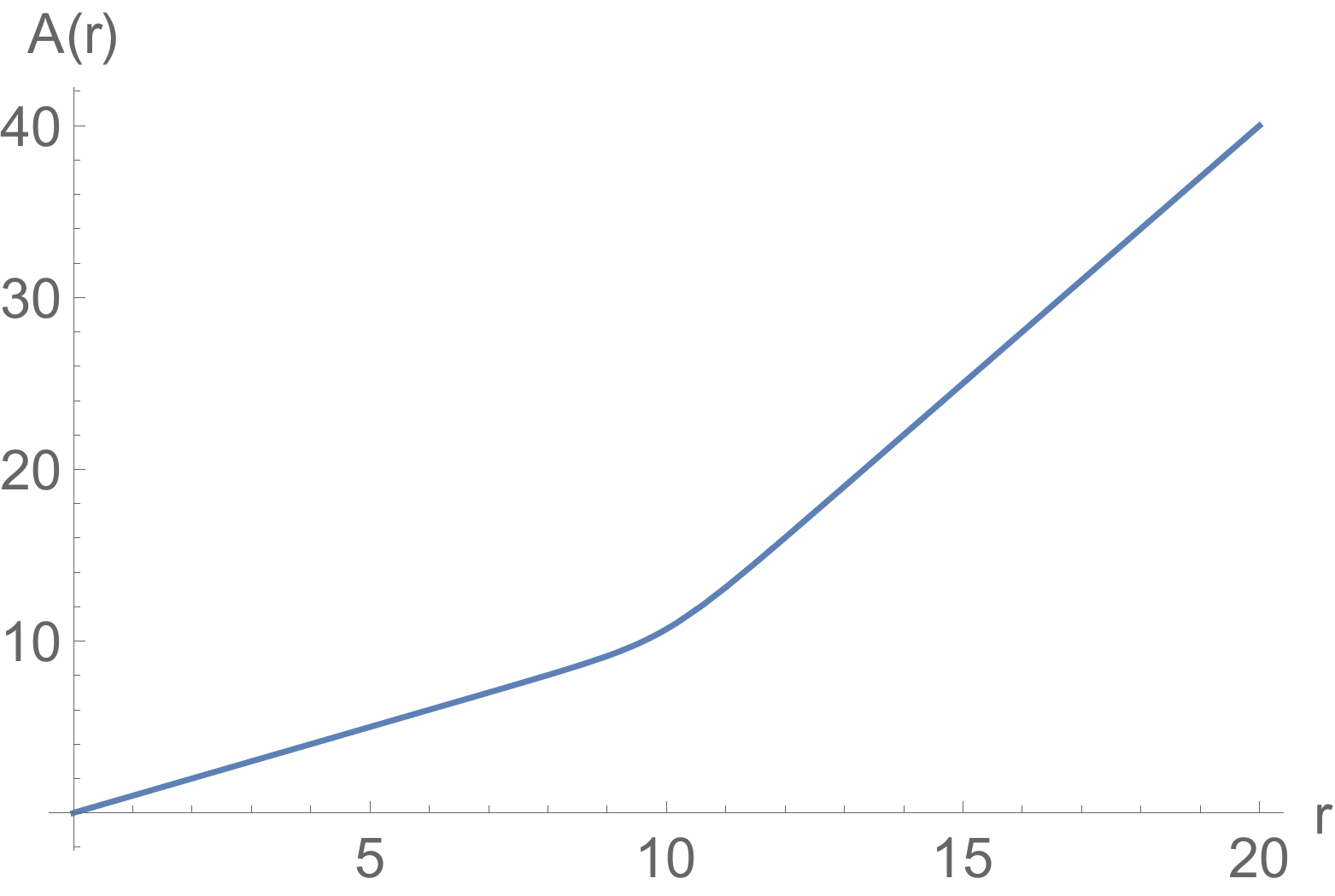}  }
		\subfigure[ ]
		{\includegraphics[width=4.8cm]{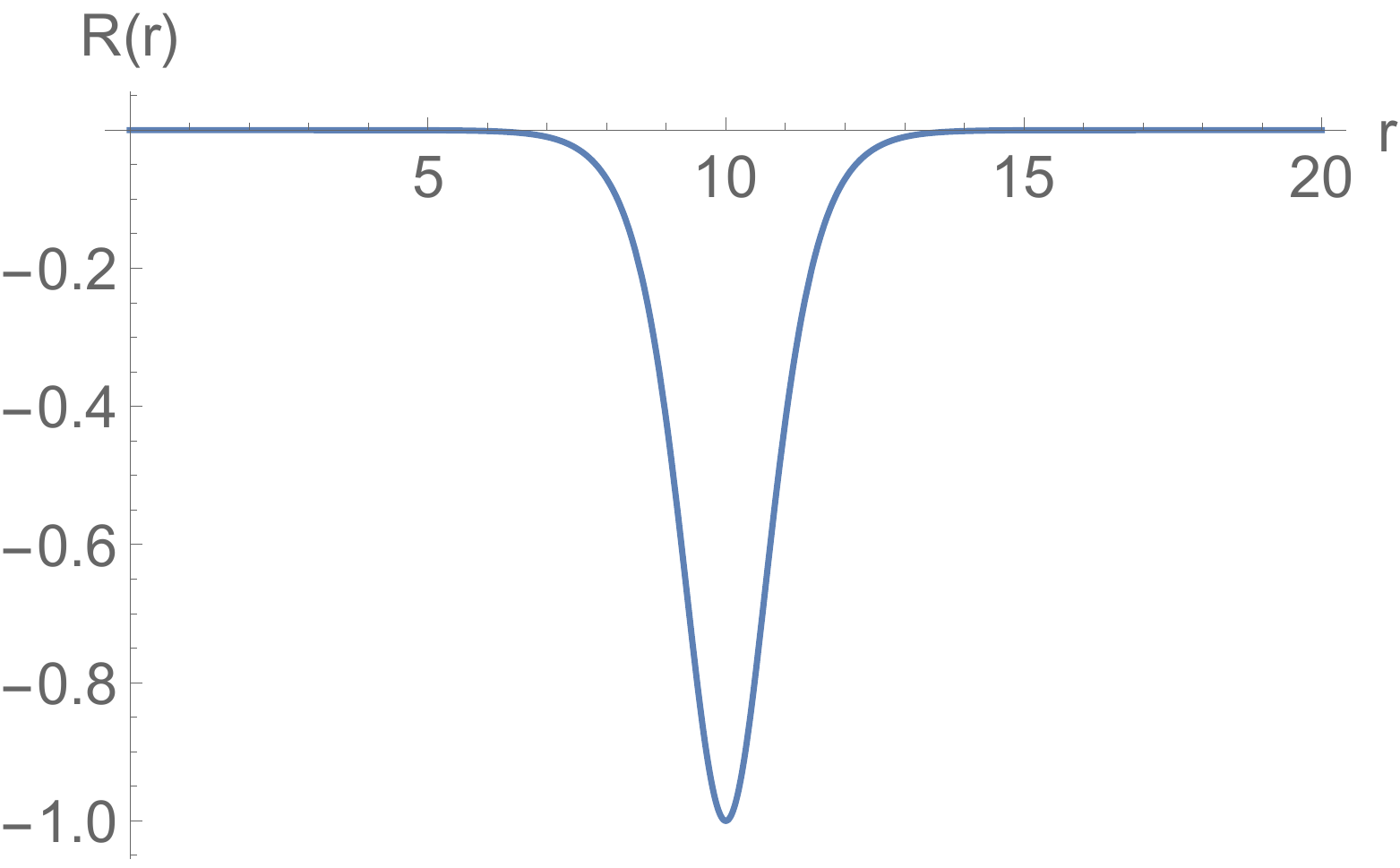}  } 
		\subfigure[ ]
		{\includegraphics[width=4.8cm]{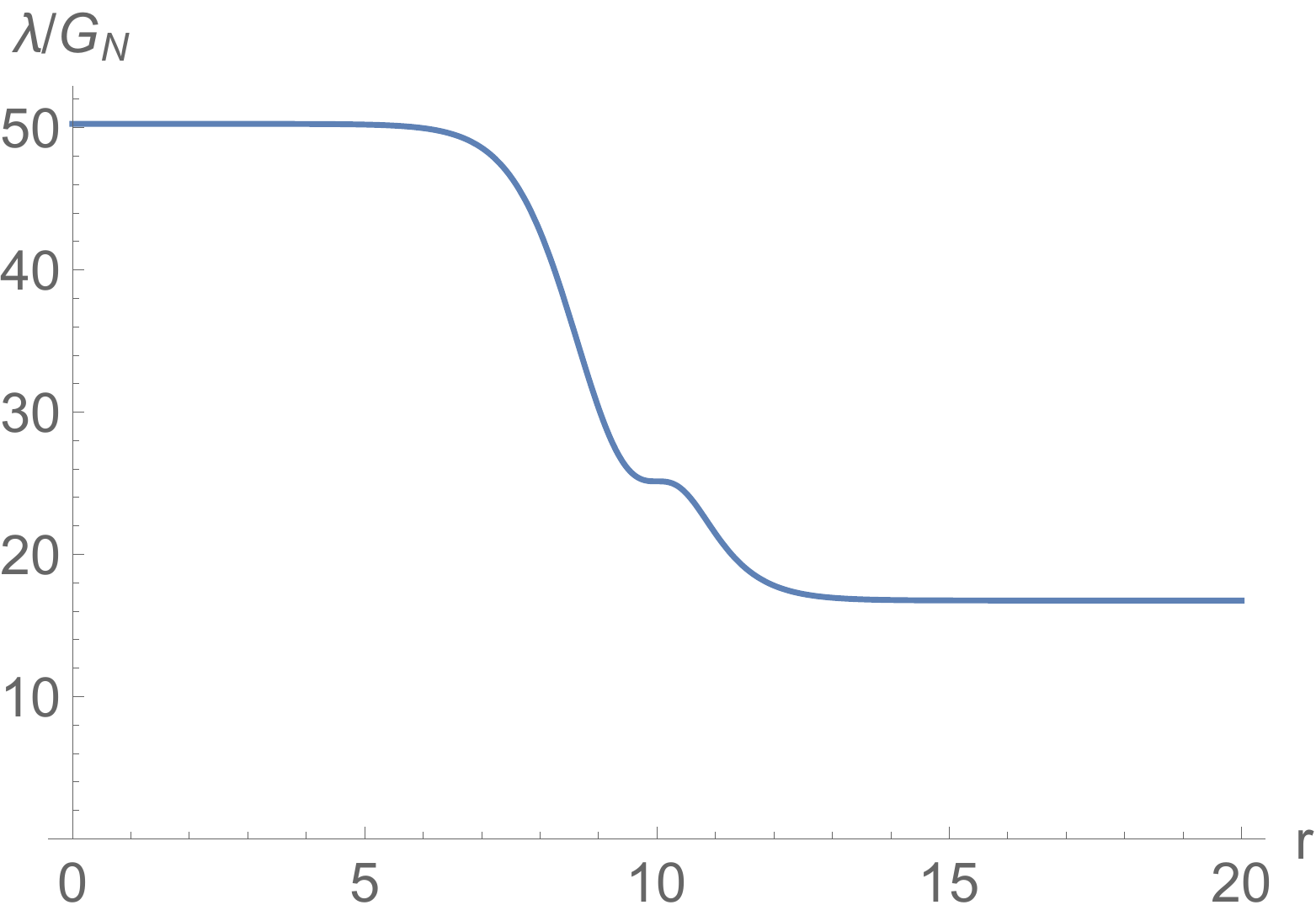}  }
		
		\caption{  Plots of $A(r)$, $R(r)$, and $\l/G_N$ for $\mathbb{A}=1, \mathbb{B}=2$ and $x_0=-10$.
		} \label{fig:AR_lambda_flat}
	\end{centering}
\end{figure}

Now, we would like to discuss another case, where the potential interpolates two flat space regions. As a representative example, we consider the following potential
\begin{align}\label{flatTwo}
W(\phi) = \mathbb{A}\tanh\phi + \mathbb{B}~,
\end{align} 
where $\mathbb{A}$ and $\mathbb{B}$ are supposed to be positive and $\mathbb{B}>\mathbb{A}$. Note that this potential has asymptotic values, $\mathbb{B}+\mathbb{A}$ and $\mathbb{B}-\mathbb{A}$ for $\phi=\infty$ and $\phi=-\infty$, respectively. When $\mathbb{A}=\mathbb{B}$, the coupling function $\lambda$ diverges. This divergent coupling can't be avoided by any field redefinition, so we consider $\mathbb{B}>\mathbb{A}$ case. Since the metric function is again given by the integration $\int_{x_0}^x d\phi\, W(\phi)$, the radial coordinate is defined by $r=(x-x_0)>0$, where $x_0$ is some small negative value chosen so that the model interpolates two regions of constant coupling. To avoid conical singularity at $r=0$, we impose a regularity condition as $\mathbb{B}-\mathbb{A}=2$. Then, the geometry is regular and flat at $r=0$. We show an example for this in Figure \ref{fig:AR_lambda_flat} with relevant parameters. The topology of the geometry is just a disc and there are two flat regions. The coupling is again a local function and has two plateau regions. Thus, when $r$ is either small or large, this model (\ref{flatTwo}) coupled to a conformal matter corresponds to a $T\bar{T}$ deformation in flat space but with different coupling constants.

%It would be interesting to study these models by considering quantum and classical fluctuations of conformal matters. Also, one can expect that topology of the spaces and the dynamical coordinate $X^a$ play important roles in associated physics. For the black hole geometries, the physical interpretation in terms of $T\bar{T}$ deformation is mysterious. We will clarify these issues in a future project.   

%%%%%%%%%%%%%%%%%%%%%%%
\section{Conclusion} 
%%%%%%%%%%%%%%%%%%%%%%%%

In this work, we have explored the effects of phase transitions of black holes in generalized JT gravity. Specifically, we considered the potential which connects two distinct ordinary JT gravities. Around the critical temperature, the metric function changes discontinuously. Geodesics and two-point functions are investigated in this context. When the potential has a locally negative region, several notable phenomena are observed: Below the critical temperature, the metric function attains a local minimum, the geodesic cannot probe inside a certain scale, and the two-point function exhibits peaks around the scale. 
We also considered charged black holes and find that the similar analysis can be done with the effective potential which includes the contribution from the gauge field. 

When the generalized JT gravity is coupled to matter, it can be related to the $T\bar{T}$ deformation of the matter action. Using the first order formalism, the action is rewritten as a $T\bar{T}$ deformation with the deforming parameter given by a function of the dilaton. The covariant derivative of the rescaled Lagrange multiplier can be interpreted as a deformed zweibein, where the recaling factor is related to the deformation parameter. Since this parameter vanishes in ordinary JT gravity, we have considered AP model instead. This corresponds to shifting the dilaton by a constant. This model is equivalent to ordinary JT gravity up to a total derivative and yields the constant deformation parameter as in flat JT gravity.   
Another example is considered where the potential interpolates two flat regions of space. This case corresponds to a Janus-type $T\bar{T}$ deformation for which the deformation parameter changes from one constant value to another.    
 
For open questions, we note that the partition function of JT gravity is equivalent to a certain matrix model integral \cite{Saad:2019lba, Stanford:2019vob}. It would be interesting to observe the phase transition of generalized JT gravity as a change in the density of states in the matrix integral formalism \cite{Maxfield:2020ale,Witten:2020wvy, Johnson:2020lns}.  

Our discussion of the $T\bar{T}$ deformation is at the classical level. For further study, computation of deformed energy spectrum would be desirable \cite{Brennan:2020dkw}. It would be interesting to study gravitational perturbations and perform a path integral analysis as in \cite{Dubovsky:2018bmo, Aguilera-Damia:2019tpe, Ishii:2019uwk, Okumura:2020dzb, Belin:2020oib}.

\acknowledgments
This work is supported by Basic Science Research Program through NRF grant No. NRF-2019R1I1A1A01057998(Y.Seo), NRF-2019R1A2C1007396(K.K.Kim, J.H. Baek). K.K.Kim acknowledges the hospitality at APCTP where part of this work was done.

\appendix

\section{Analytic Expressions }

The metric of the domain wall solution for thepotential (\ref{potential01}) is $ds^2 = A(x)d\tau^2 + \frac{dx^2}{A(x)}$ and the exact expression of $A(x)$ is given by 
\begin{align}\label{DomainW}
A(x) =& 6 s^2+50 s-\frac{5}{2} \text{Li}_2\left(-e^{2 (x+s-10)}\right)-\frac{5}{2} \text{Li}_2\left(-e^{-2 (x+s+10)}\right)+5 \text{Li}_2\left(-\frac{1}{e^{20}}\right)\nonumber\\&+6 x^2+12 s x-5 x \log \left(e^{2 (s+x-10)}+1\right)+5 x \log \left(e^{-2 (s+x+10)}+1\right)\nonumber\\&-5 s \log \left(e^{2 (s+x-10)}+1\right)-50 \log \left(e^{2 (s+x-10)}+1\right)+5 s \log \left(e^{-2 (s+x+10)}+1\right)\nonumber\\&+50 \log \left(e^{-2 (s+x+10)}+1\right)+50 \log \left(e^{2 (s+x+10)}+e^{40}\right)-50 \log (\cosh (s+x+10))\nonumber\\&+50 x-50 \left(20+\log \left(1+e^{20}\right)\right)+50 \log (\cosh (10))~,
\end{align}
where $x$ ranges from $-s$ to $\infty$. As $x$ goes to $-s$ and $\infty$, the geometry becomes two (Euclidean) $AdS_2$'s with different cosmological constants.

%%%%%%%%%%%%%%%%%%%%%%%%%%%%%%%%%%%%%%%%%%%%%%
\providecommand{\href}[2]{#2}\begingroup\raggedright

\endgroup

\providecommand{\href}[2]{#2}\begingroup\raggedright
\bibliography{references}

\begin{thebibliography}{10}

%%%%%%%CMP%%%%%%%%%%%%%%%%%%%%%%


\bibitem{Jackiw:1984je}
R.~Jackiw,
``Lower Dimensional Gravity,''
Nucl. Phys. B \textbf{252}, 343-356 (1985)
%doi:10.1016/0550-3213(85)90448-1
%543 citations counted in INSPIRE as of 22 Dec 2020
%\cite{Sachdev:1992fk}


%\cite{Teitelboim:1983ux}
\bibitem{Teitelboim:1983ux}
C.~Teitelboim,
``Gravitation and Hamiltonian Structure in Two Space-Time Dimensions,''
Phys. Lett. B \textbf{126}, 41-45 (1983)
%doi:10.1016/0370-2693(83)90012-6
%531 citations counted in INSPIRE as of 22 Dec 2020
%\cite{Jackiw:1984je}


\bibitem{Sachdev:1992fk}
S.~Sachdev and J.~Ye,
``Gapless spin fluid ground state in a random, quantum Heisenberg magnet,''
Phys. Rev. Lett. \textbf{70}, 3339 (1993)
%doi:10.1103/PhysRevLett.70.3339
\href{http://arxiv.org/abs/cond-mat/9212030}{[arXiv:cond-mat/9212030 [cond-mat]]}.


\bibitem{KitaevTalks}
A. Kitaev, Talk given at the Fundamental Physics Prize Symposium, \href{https://www.youtube.com/watch?v=OQ9qN8j7EZI}{Nov. 10, 2014}; A. Kitaev, KITP
seminar, \href{http://online.kitp.ucsb.edu/online/joint98/kitaev/}{Feb. 12, 2015}; ``A simple model of quantum holography,'' talks at KITP,  \href{http://online.kitp.ucsb.edu/online/entangled15/kitaev/}{April 7, 2015} and \href{http://online.kitp.ucsb.edu/online/entangled15/kitaev2/}{May 27, 2015}.

%\cite{Jensen:2016pah}
\bibitem{Jensen:2016pah}
K.~Jensen,
``Chaos in AdS$_2$ Holography,''
Phys. Rev. Lett. \textbf{117}, no.11, 111601 (2016)
%doi:10.1103/PhysRevLett.117.111601
\href{http://arxiv.org/abs/1605.06098}{[arXiv:1605.06098 [hep-th]]}.
%377 citations counted in INSPIRE as of 04 Dec 2020

%\cite{Maldacena:2016upp}
\bibitem{Maldacena:2016upp}
J.~Maldacena, D.~Stanford and Z.~Yang,
``Conformal symmetry and its breaking in two dimensional Nearly Anti-de-Sitter space,''
PTEP \textbf{2016}, no.12, 12C104 (2016)
%doi:10.1093/ptep/ptw124
\href{http://arxiv.org/abs/1606.01857}{[arXiv:1606.01857 [hep-th]]}.
%489 citations counted in INSPIRE as of 04 Dec 2020

%\cite{Engelsoy:2016xyb}
\bibitem{Engelsoy:2016xyb}
J.~Engels\"oy, T.~G.~Mertens and H.~Verlinde,
``An investigation of AdS$_{2}$ backreaction and holography,''
JHEP \textbf{07}, 139 (2016)
%doi:10.1007/JHEP07(2016)139
\href{http://arxiv.org/abs/1606.03438}{[arXiv:1606.03438 [hep-th]]}.
%305 citations counted in INSPIRE as of 04 Dec 2020

%\cite{Almheiri:2019hni}
\bibitem{Almheiri:2019hni}
A.~Almheiri, R.~Mahajan, J.~Maldacena and Y.~Zhao,
``The Page curve of Hawking radiation from semiclassical geometry,''
JHEP \textbf{03}, 149 (2020)
%doi:10.1007/JHEP03(2020)149
\href{http://arxiv.org/abs/1908.10996}{[arXiv:1908.10996 [hep-th]]}.
%127 citations counted in INSPIRE as of 03 Dec 2020

%\cite{Almheiri:2019qdq}
\bibitem{Almheiri:2019qdq}
A.~Almheiri, T.~Hartman, J.~Maldacena, E.~Shaghoulian and A.~Tajdini,
``Replica Wormholes and the Entropy of Hawking Radiation,''
JHEP \textbf{05}, 013 (2020)
%doi:10.1007/JHEP05(2020)013
\href{http://arxiv.org/abs/1911.12333}{[arXiv:1911.12333 [hep-th]]}.
%143 citations counted in INSPIRE as of 03 Dec 2020

%\cite{Witten:2020ert}
\bibitem{Witten:2020ert}
E.~Witten,
``Deformations of JT Gravity and Phase Transitions,''
\href{http://arxiv.org/abs/2006.03494}{[arXiv:2006.03494 [hep-th]]}.
%17 citations counted in INSPIRE as of 27 Nov 2020




%\cite{Smirnov:2016lqw}
\bibitem{Smirnov:2016lqw}
F.~A.~Smirnov and A.~B.~Zamolodchikov,
``On space of integrable quantum field theories,''
Nucl. Phys. B \textbf{915}, 363-383 (2017)
%doi:10.1016/j.nuclphysb.2016.12.014
\href{http://arxiv.org/abs/1608.05499}{[arXiv:1608.05499 [hep-th]]}.
%197 citations counted in INSPIRE as of 10 Dec 2020

%\cite{Cavaglia:2016oda}
\bibitem{Cavaglia:2016oda}
A.~Cavagli\`a, S.~Negro, I.~M.~Sz\'ecs\'enyi and R.~Tateo,
``$T \bar{T}$-deformed 2D Quantum Field Theories,''
JHEP \textbf{10}, 112 (2016)
%doi:10.1007/JHEP10(2016)112
\href{http://arxiv.org/abs/1608.05534}{[arXiv:1608.05534 [hep-th]]}.
%179 citations counted in INSPIRE as of 10 Dec 2020

%\cite{Dubovsky:2018bmo}
\bibitem{Dubovsky:2018bmo}
S.~Dubovsky, V.~Gorbenko and G.~Hern\'andez-Chifflet,
``$ T\overline{T} $ partition function from topological gravity,''
JHEP \textbf{09}, 158 (2018)
%doi:10.1007/JHEP09(2018)158
\href{http://arxiv.org/abs/1805.07386}{[arXiv:1805.07386 [hep-th]]}.
%90 citations counted in INSPIRE as of 27 Nov 2020

%\cite{Dubovsky:2017cnj}
\bibitem{Dubovsky:2017cnj}
S.~Dubovsky, V.~Gorbenko and M.~Mirbabayi,
``Asymptotic fragility, near AdS$_{2}$ holography and $ T\overline{T} $,''
JHEP \textbf{09}, 136 (2017)
%doi:10.1007/JHEP09(2017)136
\href{http://arxiv.org/abs/1706.06604}{[arXiv:1706.06604 [hep-th]]}.
%118 citations counted in INSPIRE as of 17 Dec 2020

%\cite{Cardy:2018sdv}
\bibitem{Cardy:2018sdv}
J.~Cardy,
``The $ T\overline{T} $ deformation of quantum field theory as random geometry,''
JHEP \textbf{10}, 186 (2018)
%doi:10.1007/JHEP10(2018)186
\href{http://arxiv.org/abs/1801.06895}{[arXiv:1801.06895 [hep-th]]}.
%106 citations counted in INSPIRE as of 17 Dec 2020



%\cite{Conti:2018tca}
\bibitem{Conti:2018tca}
R.~Conti, S.~Negro and R.~Tateo,
``The $ \mathrm{T}\overline{\mathrm{T}} $ perturbation and its geometric interpretation,''
JHEP \textbf{02}, 085 (2019)
%doi:10.1007/JHEP02(2019)085
\href{http://arxiv.org/abs/1809.09593}{[arXiv:1809.09593 [hep-th]]}.
%49 citations counted in INSPIRE as of 09 Dec 2020

%\cite{Conti:2019dxg}
\bibitem{Conti:2019dxg}
R.~Conti, S.~Negro and R.~Tateo,
``Conserved currents and $\text{T}\bar{\text{T}}_s$ irrelevant deformations of 2D integrable field theories,''
JHEP \textbf{11}, 120 (2019)
%doi:10.1007/JHEP11(2019)120
\href{http://arxiv.org/abs/1904.09141}{[arXiv:1904.09141 [hep-th]]}.
%32 citations counted in INSPIRE as of 09 Dec 2020

%\cite{Coleman:2019dvf}
\bibitem{Coleman:2019dvf}
E.~A.~Coleman, J.~Aguilera-Damia, D.~Z.~Freedman and R.~M.~Soni,
``$ T\overline{T} $ -deformed actions and (1,1) supersymmetry,''
JHEP \textbf{10}, 080 (2019)
%doi:10.1007/JHEP10(2019)080
\href{http://arxiv.org/abs/1906.05439}{[arXiv:1906.05439 [hep-th]]}.
%23 citations counted in INSPIRE as of 09 Dec 2020

%\cite{Aguilera-Damia:2019tpe}
\bibitem{Aguilera-Damia:2019tpe}
J.~Aguilera-Damia, V.~I.~Giraldo-Rivera, E.~A.~Mazenc, I.~Salazar Landea and R.~M.~Soni,
``A path integral realization of joint $ J\overline{T} $, $ T\overline{J} $ and $ T\overline{T} $ flows,''
JHEP \textbf{07}, no.07, 085 (2020)
%doi:10.1007/JHEP07(2020)085
\href{http://arxiv.org/abs/1910.06675}{[arXiv:1910.06675 [hep-th]]}.
%9 citations counted in INSPIRE as of 09 Dec 2020

%\cite{Tolley:2019nmm}
\bibitem{Tolley:2019nmm}
A.~J.~Tolley,
``$ T\overline{T} $ deformations, massive gravity and non-critical strings,''
JHEP \textbf{06}, 050 (2020)
%doi:10.1007/JHEP06(2020)050
\href{http://arxiv.org/abs/1911.06142}{[arXiv:1911.06142 [hep-th]]}.
%19 citations counted in INSPIRE as of 09 Dec 2020

%\cite{Mazenc:2019cfg}
\bibitem{Mazenc:2019cfg}
E.~A.~Mazenc, V.~Shyam and R.~M.~Soni,
``A $T \bar{T}$ Deformation for Curved Spacetimes from 3d Gravity,''
\href{http://arxiv.org/abs/1912.09179}{[arXiv:1912.09179 [hep-th]]}.
%15 citations counted in INSPIRE as of 09 Dec 2020

%\cite{Caputa:2020lpa}
\bibitem{Caputa:2020lpa}
P.~Caputa, S.~Datta, Y.~Jiang and P.~Kraus,
``Geometrizing $T\bar{T}$,''
\href{http://arxiv.org/abs/2011.04664}{[arXiv:2011.04664 [hep-th]]}.
%1 citations counted in INSPIRE as of 09 Dec 2020

%\cite{Almheiri:2014cka}
\bibitem{Almheiri:2014cka}
A.~Almheiri and J.~Polchinski,
``Models of AdS$_{2}$ backreaction and holography,''
JHEP \textbf{11}, 014 (2015)
%doi:10.1007/JHEP11(2015)014
\href{http://arxiv.org/abs/1402.6334}{[arXiv:1402.6334 [hep-th]]}.
%325 citations counted in INSPIRE as of 01 Dec 2020

%\cite{Brennan:2020dkw}
\bibitem{Brennan:2020dkw}
T.~D.~Brennan, C.~Ferko, E.~Martinec and S.~Sethi,
``Defining the $T \overline{T}$ Deformation on $\mathrm{AdS}_2$,''
\href{http://arxiv.org/abs/2005.00431}{[arXiv:2005.00431 [hep-th]]}.
%6 citations counted in INSPIRE as of 09 Dec 2020





%%%%%%%%%%% end of introduction %%%%%%%%%%%%%%%%%%%%%%%%%%%%%%%%%%%%%%%%%





%\cite{Strominger:1998yg}
\bibitem{Strominger:1998yg}
A.~Strominger,
%``AdS(2) quantum gravity and string theory,''
JHEP \textbf{01}, 007 (1999)
%doi:10.1088/1126-6708/1999/01/007
\href{http://arxiv.org/abs/hep-th/9809027}{[arXiv:hep-th/9809027 [hep-th]]}.
%284 citations counted in INSPIRE as of 17 Dec 2020

%\cite{Hartman:2008dq}
\bibitem{Hartman:2008dq}
T.~Hartman and A.~Strominger,
``Central Charge for AdS(2) Quantum Gravity,''
JHEP \textbf{04}, 026 (2009)
%doi:10.1088/1126-6708/2009/04/026
\href{http://arxiv.org/abs/0803.3621}{[arXiv:0803.3621 [hep-th]]}.
%93 citations counted in INSPIRE as of 17 Dec 2020

%\cite{Castro:2008ms}
\bibitem{Castro:2008ms}
A.~Castro, D.~Grumiller, F.~Larsen and R.~McNees,
``Holographic Description of AdS(2) Black Holes,''
JHEP \textbf{11}, 052 (2008)
%doi:10.1088/1126-6708/2008/11/052
\href{http://arxiv.org/abs/0809.4264}{[arXiv:0809.4264 [hep-th]]}.
%100 citations counted in INSPIRE as of 17 Dec 2020

%\cite{Cvetic:2016eiv}
\bibitem{Cvetic:2016eiv}
M.~Cveti\v{c} and I.~Papadimitriou,
``AdS$_{2}$ holographic dictionary,''
JHEP \textbf{12}, 008 (2016)
[erratum: JHEP \textbf{01}, 120 (2017)]
%doi:10.1007/JHEP12(2016)008
\href{http://arxiv.org/abs/1608.07018}{[arXiv:1608.07018 [hep-th]]}.
%123 citations counted in INSPIRE as of 17 Dec 2020

%\cite{Lala:2019inz}
\bibitem{Lala:2019inz}
A.~Lala and D.~Roychowdhury,
``Models of phase stability in Jackiw-Teitelboim gravity,''
Phys. Rev. D \textbf{100}, 124061 (2019)
%doi:10.1103/PhysRevD.100.124061
\href{http://arxiv.org/abs/1909.09828}{[arXiv:1909.09828 [hep-th]]}.
%3 citations counted in INSPIRE as of 24 Dec 2020

%\cite{Lala:2020lge}
\bibitem{Lala:2020lge}
A.~Lala, H.~Rathi and D.~Roychowdhury,
%``Jackiw-Teitelboim gravity and the models of a Hawking-Page transition for 2D black holes,''
Phys. Rev. D \textbf{102}, no.10, 104024 (2020)
%doi:10.1103/PhysRevD.102.104024
\href{http://arxiv.org/abs/2005.08018}{[arXiv:2005.08018 [hep-th]]}.
%2 citations counted in INSPIRE as of 24 Dec 2020

%\cite{Saad:2019lba}
\bibitem{Saad:2019lba}
P.~Saad, S.~H.~Shenker and D.~Stanford,
``JT gravity as a matrix integral,''
\href{http://arxiv.org/abs/1903.11115}{[arXiv:1903.11115 [hep-th]]}.
%173 citations counted in INSPIRE as of 03 Dec 2020

%\cite{Stanford:2019vob}
\bibitem{Stanford:2019vob}
D.~Stanford and E.~Witten,
``JT Gravity and the Ensembles of Random Matrix Theory,''
\href{http://arxiv.org/abs/1907.03363}{[arXiv:1907.03363 [hep-th]]}.
%94 citations counted in INSPIRE as of 17 Dec 2020


%\cite{Maxfield:2020ale}
\bibitem{Maxfield:2020ale}
H.~Maxfield and G.~J.~Turiaci,
``The path integral of 3D gravity near extremality; or, JT gravity with defects as a matrix integral,''
\href{http://arxiv.org/abs/2006.11317}{[arXiv:2006.11317 [hep-th]]}.
%35 citations counted in INSPIRE as of 19 Jan 2021


%\cite{Witten:2020wvy}
\bibitem{Witten:2020wvy}
E.~Witten,
``Matrix Models and Deformations of JT Gravity,''
\href{http://arxiv.org/abs/2006.13414}{[arXiv:2006.13414 [hep-th]]}.
%22 citations counted in INSPIRE as of 27 Nov 2020


%\cite{Johnson:2020lns}
\bibitem{Johnson:2020lns}
C.~V.~Johnson and F.~Rosso,
``Solving Puzzles in Deformed JT Gravity: Phase Transitions and Non-Perturbative Effects,''
\href{http://arxiv.org/abs/2011.06026}{[arXiv:2011.06026 [hep-th]]}.
%1 citations counted in INSPIRE as of 24 Dec 2020 


%\cite{Ishii:2019uwk}
\bibitem{Ishii:2019uwk}
T.~Ishii, S.~Okumura, J.~I.~Sakamoto and K.~Yoshida,
``Gravitational perturbations as $T\bar{T}$-deformations in 2D dilaton gravity systems,''
Nucl. Phys. B \textbf{951}, 114901 (2020)
%doi:10.1016/j.nuclphysb.2019.114901
\href{http://arxiv.org/abs/1906.03865}{[arXiv:1906.03865 [hep-th]]}.
%10 citations counted in INSPIRE as of 17 Dec 2020
 
%\cite{Okumura:2020dzb}
\bibitem{Okumura:2020dzb}
S.~Okumura and K.~Yoshida,
``$T\bar{T}$-deformation and Liouville gravity,''
Nucl. Phys. B \textbf{957}, 115083 (2020)
%doi:10.1016/j.nuclphysb.2020.115083
\href{http://arxiv.org/abs/2003.14148}{[arXiv:2003.14148 [hep-th]]}.
%3 citations counted in INSPIRE as of 17 Dec 2020 

%\cite{Belin:2020oib}
\bibitem{Belin:2020oib}
A.~Belin, A.~Lewkowycz and G.~Sarosi,
``Gravitational path integral from the $T^2$ deformation,''
JHEP \textbf{09}, 156 (2020)
%doi:10.1007/JHEP09(2020)156
\href{http://arxiv.org/abs/2006.01835}{[arXiv:2006.01835 [hep-th]]}.
%8 citations counted in INSPIRE as of 17 Dec 2020
 

 
 
 



%%%%%%%%%%%%%%%%%%
%%%%%%%%%%%%%%%%%%
%%%%%%%%%%%%%%%%%%


\end{thebibliography}
\bibliographystyle{JHEP}
\endgroup

\end{document}